%% file: bhpairs_tremmel_2017b_v3_complete.tex
\documentclass[useAMS,usenatbib,twocolumn]{mn2e}
\usepackage{natbib,graphics,epsfig}
\usepackage{amsmath}
\usepackage{amssymb}
\usepackage{textcomp}
\usepackage{graphicx}
\usepackage{color}
\usepackage{hyperref}
\usepackage{txfonts}
\input{rgb}

\usepackage{graphicx}
\usepackage{color}
\input{rgb}
\usepackage[T1]{fontenc}
\usepackage{aecompl}

\begin{document}

\title[Dancing to ChaNGa]{Dancing to ChaNGa:  A Self-Consistent Prediction For Close SMBH Pair Formation Timescales Following Galaxy Mergers}
\author[M. Tremmel et al.]{M. ~Tremmel$^{1,2}$\thanks{email: michael.tremmel@yale.edu}, F. ~Governato$^{1}$, M. 
~Volonteri $^{3}$, T. ~R. ~Quinn$^{1}$, A. ~Pontzen$^{4}$\\
$^1$Astronomy Department, University of Washington, Box 351580, Seattle, WA, 98195-1580\\
$^2$Yale Center for Astronomy \& Astrophysics, Physics Department, P.O. Box 208120, New Haven, CT 06520, USA\\
$^3$Sorbonne Universit\`{e}s, UPMC Univ Paris 6 et CNRS, UMR 7095, Institut d`Astrophysique de Paris, 98 bis bd Arago, 75014 Paris, France\\
$^4$ Department of Physics and Astronomy, University College London, 132 Hampstead Road, London, NW1 2PS, 
United Kingdom}

\pagerange{\pageref{firstpage}--\pageref{lastpage}} \pubyear{2015}

\maketitle

\label{firstpage}

\begin{abstract}
We present the first self-consistent prediction for the distribution of formation timescales for close Supermassive Black Hole (SMBH) pairs following galaxy mergers. Using {\sc Romulus25}, the first large-scale cosmological simulation to accurately track the orbital evolution of SMBHs within their host galaxies down to sub-kpc scales, we predict an average formation rate density of close SMBH pairs of 0.013 cMpc$^{-3}$ Gyr$^{-1}$. We find that it is relatively rare for galaxy mergers to result in the formation of close SMBH pairs with sub-kpc separation and those that do form are often the result of Gyrs of orbital evolution following the galaxy merger. The likelihood and timescale to form a close SMBH pair depends strongly on the mass ratio of the merging galaxies, as well as the presence of dense stellar cores. Low stellar mass ratio mergers with galaxies that lack a dense stellar core are more likely to become tidally disrupted and deposit their SMBH at large radii without any stellar core to aid in their orbital decay, resulting in a population of long-lived `wandering' SMBHs. Conversely, SMBHs in galaxies that remain embedded within a stellar core form close pairs in much shorter timescales   on average.  This timescale is a crucial, though often ignored or very simplified, ingredient to models predicting SMBH mergers rates and the connection between SMBH and star formation activity.

\end{abstract}



\section{Introduction}

 Despite their importance to galaxy evolution theory, the mechanisms driving the co-evolution of supermassive black holes (SMBHs) and their host galaxies, and indeed the processes that form SMBHs in the first place, are highly uncertain.
SMBHs are ubiquitous in galaxies ranging from massive ellipticals and bulge-dominated galaxies \citep[e.g.][]{Gehren84,kormendy95,kormendy2013} to smaller, bulge-less disk galaxies and dwarfs \citep{shields08,Filippenko03,reines11,reines12,reines13,moran14}. Empirical scaling relationships between the mass of SMBHs and that of their host galaxies are indicative of coeval growth \citep{haring04,gultekin09, schramm2013, kormendy2013,volonteribellovary12}.

Future observations of gravitational waves emitted from binary and merging SMBHs via pulsar timing arrays \citep{sesana13} and the planned LISA mission \citep{klein16} will provide unique information on the SMBH population and its dynamical evolution. Pulsar timing arrays probe relatively low-redshift (z<~2) BHs towards the high mass end ($> 10^8$ M$_{\odot}$), while LISA can detect mergers of SMBHs with mass $\sim10^4-10^7$ M$_{\odot}$ out to the highest redshift. LISA has therefore the capability to  provide unique constraints to the SMBH mass function across cosmic time as well as critical insight into their possible formation mechanisms \citep{sesana07,volonteriMSIGMA2009,klein16} and their growth and spin evolution \citep{Berti08,Barausse12}. Further, on-going observations, as well as large-scale cosmological simulations, of active SMBHs that are offset from the centre of their host galaxies, possibly in galaxies with multiple luminous SMBHs, can potentially help constrain the extent to which galaxy mergers drive SMBH growth \citep{comerford15, steinborn15DualAGN,barrows17}. 

The formation of a SMBH binary and subsequent merger of two SMBHs can be described in a number of stages. First, a dark matter halo falls into a halo of larger mass. It then sinks to the centre via dynamical friction and the two central galaxies then begin to strongly interact and merge. Following the merger of two galaxies hosting SMBHs, dynamical friction acting on the SMBHs causes them to sink to galactic centre and form a close pair with sub-kiloparsec (kpc) separation. The close pair, through dynamical interactions with gas and stars, then forms a bound SMBH binary ($D<10$ pc), which then itself hardens to the point where gravitational wave emission causes rapid orbital decay and the two SMBHs merge ($D<0.001$pc).

While the orbital evolution of close SMBH pairs and the bound binary systems that follow are extensively studied using numerical and analytic techniques \citep[e.g.][]{armitage02,Yu02,sesana15,dosopoulou17}, it is also critical to understand the evolution of SMBH pairs on larger scales, as these timescales can be quite long \citep[e.g.][]{callegariBH09,callegari11} and present a critical bottleneck to SMBH binary formation. However, studies of SMBH orbital evolution prior to the formation of close pairs has so far been severely limited. Semi-analytic models account for this timescale using simple models for dynamical friction \citep[e.g.][]{dosopoulou17,dvorkin17}. Detailed simulations of isolated mergers have indicated that SMBH sinking timescales following major mergers depend on the central stellar density of both galaxies \citep{G94} and can be quite short \citep{mayer07}, while SMBH sinking timescales following minor mergers can be much longer and depend sensitively on the orientation of the merging galaxies \citep{callegariBH09,callegari11}. However, these idealized simulations do not produce the realistic merger and gas accretion histories that real galaxies experience in a full cosmological context.



Cosmological simulations potentially  provide a more self-consistent view of SMBH orbital decay timescales and are the logical next step from isolated galaxy merger simulations to better understand the timescales of close pair formation. With these simulations, the effects of different  morphology and merger dynamics are naturally accounted for without \textit{a priori} assumptions, as each galaxy in the simulation has a cosmologically realistic accretion and merger history. However, past simulations generally had poor resolution, which required simplified assumptions such as `advection', where SMBHs quickly sink into
the deepest nearby potential well, resulting in unrealistic, nearly instantaneous SMBH orbital decay. This approximation contrasts with the above numerical results as it assumes that the orbital sinking timescale on kpc scales is effectively zero. In previous works we have shown that this technique often results in inaccurate SMBH dynamics within galaxies and a drastic underestimate of sinking timescales \citep{tremmel15}. While current simulations are beginning to employ more detailed approaches to SMBH dynamics \citep[e.g.][]{Hirschmann14,dubois16,steinborn15DualAGN}, accurate orbital evolution down to sub-kpc scales remains a challenge.

In this Paper, using the {\sc Romulus25} cosmological simulation \citep{tremmel17} which is uniquely able to predict the orbital evolution of SMBHs down to sub-kpc scales \citep{tremmel15}, \textit{we present the first robust estimate of SMBH sinking and subsequent close SMBH pair formation timescales over a range of cosmic epochs and galaxy properties.}



\section{The Romulus Simulations}

The {\sc Romulus} Simulations are a set of large-scale, high resolution cosmological simulations with emphasis on implementing a novel approach to SMBH formation, dynamics, and accretion. For this work, we focus on {\sc Romulus25}, our flagship 25 Mpc per side volume simulation, as it provides a uniform sample of galaxies within a wide range of halo masses ($3\times10^{9}$ to $2\times10^{13}$ M$_{\odot}$). The simulation is run assuming a $\Lambda$CDM cosmology following the most recent results from Planck \citep[$\Omega_0=0.3086$, $\Lambda=0.6914$, h$=0.67$, $\sigma_8=0.77$;][]{planck16}, a Plummer equivalent force softening of $250$ pc (a $350$ pc spline force softening is used), and mass resolution for dark matter and gas of $3.39 \times 10^5$ and $2.12 \times 10^5$ M$_{\odot}$ respectively. The simulation was run using the new Tree + SPH code, {\sc ChaNGa} \citep{changa15}, which includes an updated SPH implementation that accurately simulates shearing flows with Kelvin-Helmholtz instabilities. The Simulations also include the standard physics modules previously used in  { \sc GASOLINE}, such as a cosmic UV background, star formation, `blastwave' SN feedback, low temperature metal cooling  \citep{wadsley04,wadsley08,Stinson06,shen10}, as well as a novel implementation of SMBH formation, growth, and dynamics \citep{tremmel15,tremmel17}.
 
 As described in more detail in \citet{tremmel17},  the free parameters within our sub-grid models for star formation and SMBH physics (see \S2.1) are optimized and held constant. This was achieved using a large set of `zoomed-in' simulations of  galaxies within dark matter halos with masses $10^{10.5}$, $10^{11.5}$, and $10^{12}$ M$_{\odot}$. Each set of galaxies was 1) run using a different set of parameters and 2)  graded against different $z = 0$ scaling relations related to star formation efficiency, gas fraction, angular momentum, and black hole growth. This resulted in {\it fully specified } sub-grid models governing star formation, stellar feedback, and SMBH accretion and feedback that are optimized to provide realistic $z = 0$ galaxies while maintaining predictive power at higher redshifts and high mass (M$_{h} > 10^{12}$ M$_{\odot}$). {\sc Romulus25} has been shown to reproduce the z$=$0 stellar mass halo mass and SMBH mass stellar mass relations across the entire range of resolved halos. It also {\it predicts} cosmic star formation and SMBH accretion histories at high redshift that are consistent with observations \citep{tremmel17}.
 
 \subsection{SMBH Accretion and Feedback}
 
Accretion of gas onto SMBHs is governed by a modified Bondi-Hoyle prescription. Using the same energy balance argument as in the derivation of Bondi-Hoyle, we re-derive the SMBH accretion radius to include the effect of angular momentum support based on the resolved dynamics of gas in the simulation. We also apply a density dependent boost factor to account for the unresolved multiphase nature of the ISM near a SMBH \citep{BoothBH2009}, giving us the final equation

\begin{equation}
\dot{M} = \left ( \frac{n}{n_{th,*}} \right )^\beta\begin{cases}
\frac{\pi(GM)^2 \rho}{(v_{\mathrm{bulk}}^2+c_s^2)^{3/2}} & \text{ if } v_{\mathrm{bulk}}>v_{\theta} \\ \\
\frac{\pi(GM)^2 \rho c_s}{(v_{\theta}^2+c_s^2)^{2}} & \text{ if }  v_{\mathrm{bulk}}<v_{\theta}.
\end{cases}
\end{equation}

\noindent The tangential velocity, $v_{\theta}$, is estimated at the smallest resolved scales in the simulation and compared to $v_{\mathrm{bulk}}$, the overall bulk motion of the gas that already enters into the Bondi-Hoyle model. When the bulk motion dominates over the nearby rotational motion, or the energetics are dominated by the internal energy of the gas, the accretion reverts to the normal Bondi-Hoyle prescription. The threshold for star formation, $n_{th,*}$, also determines the threshold beyond which we assume gas becomes multiphase and poorly resolved, requiring a boost to the approximated accretion rate. For lower densities, we assume that the gas is not sufficiently multiphase to require such a boost, as in \citet{BoothBH2009}. How much this boost increases with density is governed by $\beta$, constrained by our parameter search to be 2.

An accreting SMBH converts a fraction of that mass, $\epsilon_r$, into energy. A fraction of this energy, $\epsilon_f$, is thermally coupled to the 32 nearest gas particles according to the smoothing kernel. We assume the common value of 10\% for $\epsilon_r$ and take $\epsilon_f$ as a free parameter again set by our parameter search technique to be 0.02. For more details on SMBH accretion and feedback in {\sc Romulus}, we refer to the reader to \citet{tremmel17}. We note that while there still exists issues with the Bondi-Hoyle formalism even in the regime of non-rotating gas \citep[e.g.][]{hobbs12}, for the spatial and time resolution of these simulations, it still represents the best way of approximating long term accretion onto SMBHs based on large-scale gas properties without requiring additional assumptions.

\subsection{SMBH Seeding}


SMBHs are seeded in the simulation based on gas properties, forming in rapidly collapsing, low metallicity regions in the early Universe. We isolate pristine gas particles ($Z<3\times10^{-4}$) that have reached densities 15 times higher than what is required by our star formation prescription without forming a star or cooling beyond $9.5 \times 10^3$ K (just below the temperature threshold used for star formation, $10^{4}$ K). These regions are collapsing on timescales much shorter than the cooling and star formation timescales and are meant to approximate the regions that would form large SMBHs, regardless of the details of their formation mechanism. \citet{tremmel17} show how this method forms most SMBHs within the first Gyr of the simulation, compared with the later seeding times inherent to common approaches that seed based on halo mass thresholds \citep[e.g.][]{diMatteo2008,genel14,eagle15}.  


The seed SMBH mass is set to $10^6$ M$_{\odot}$ and is justified by our choice of formation criteria, which would produce SMBHs that are able to attain higher masses quickly, as there is a lot of dense, collapsing gas nearby that is unlikely to form stars. Critically for our analysis presented here, this initial mass guarantees that SMBHs always have a mass significantly larger than DM and gas particles, allowing us to correctly resolve their dynamics without resorting to ad hoc simplifications \citep{tremmel15}. This approach results in an evolving occupation fraction. At early times, small halos (M$_{vir} \sim 10^{9-10}$ M$_{\odot}$) host massive, newly formed SMBHs. The occupation fraction evolves due to hierarchical merging and the fact that halos in less dense regions are less likely to host such dense collapsing regions at early times. Less than 10\% of halos with $3\times 10^9 <$ M$_{vir} < 10^{10}$ M$_{\odot}$ host a SMBH of mass at least $10^6$ M$_{\odot}$ at $z = 0$. Beyond the scope of this study is the examination of less massive black holes more common in smaller halos \citep[e.g.][]{ReinesVolonteri15,baldassare16}. Their lower masses will make them less likely to sink to the centre of their new host halo following a galaxy merger.

\subsection{SMBH Dynamics}

Following the merger of two galaxies hosting SMBHs, the accreted SMBHs sink toward the centre of the descendant galaxy through dynamical friction, the force exerted by the gravitational wake caused by a massive body moving through an extended medium \citep{DF43,stelios05,BinneyTremaine}. However, the limited mass and gravitational force resolution of cosmological simulations leaves this process largely unresolved. The Romulus simulations uniquely include the sub-grid correction accounting for this unresolved dynamical friction described in \citet{tremmel15} that has been shown to produce realistically sinking SMBHs (see appendix A for tests of this prescription at the specific resolution of {\sc Romulus25}).

As described in detail in \citet{tremmel15}, the dynamical friction acting on a SMBH of mass M from surrounding star and dark matter (DM) particles is approximated using Chandrasekhar's dynamical friction formula \citep{DF43} integrated out to our softening limit, $\epsilon_{g}$, and assuming a locally isotropic velocity distribution and constant density out to $\epsilon_{g}$ .

\begin{equation}
\mathbf{a}_{DF} = -4\pi\mathrm{G}^2\mathrm{M}\rho(<\mathrm{v}_{BH})\mathrm{ln}\Lambda\frac{\mathbf{v}_{BH}}{\mathrm{v}_{BH}^3}.
\end{equation}

\noindent The velocity of the SMBH, $\mathrm{v}_{BH}$, is taken relative to the local centre of mass (COM) velocity of stars and DM. We have also assumed that the contribution from objects moving faster than the SMBH is negligible, where $\rho(<\mathrm{v}_{BH}$) is the density of particles moving slower than the SMBH relative to the local COM. This is a good assumption to make for dynamical evolution on scales much larger than 1 pc \citep{antonini12}. The coulomb logarithm, $\mathrm{ln}\Lambda$, is taken to be $\mathrm{ln}(\frac{\mathrm{b}_{max}}{\mathrm{b}_{min}})$, where b$_{max}$ = $\epsilon_{g}$ to avoid double counting the resolved dynamical friction that is already occurring on larger scales and  b$_{min}$ is the $90^{\circ}$ deflection radius, with a lower limit set to the Schwarzschild Radius, $\mathrm{R}_{Sch}$. The calculation is done based the 64 nearest star and DM particles to each SMBH. {\sc Romulus25} achieves mass resolution such that the ambient dark matter, gas, and star particles are significantly less massive than the smallest SMBHs, allowing it to avoid the numerical effects that persist at low resolution even with this dynamical friction prescription \citep{tremmel15}.

 \begin{figure}
\centering
\includegraphics[trim=10mm 5mm -10mm 0mm, clip, width=90mm]{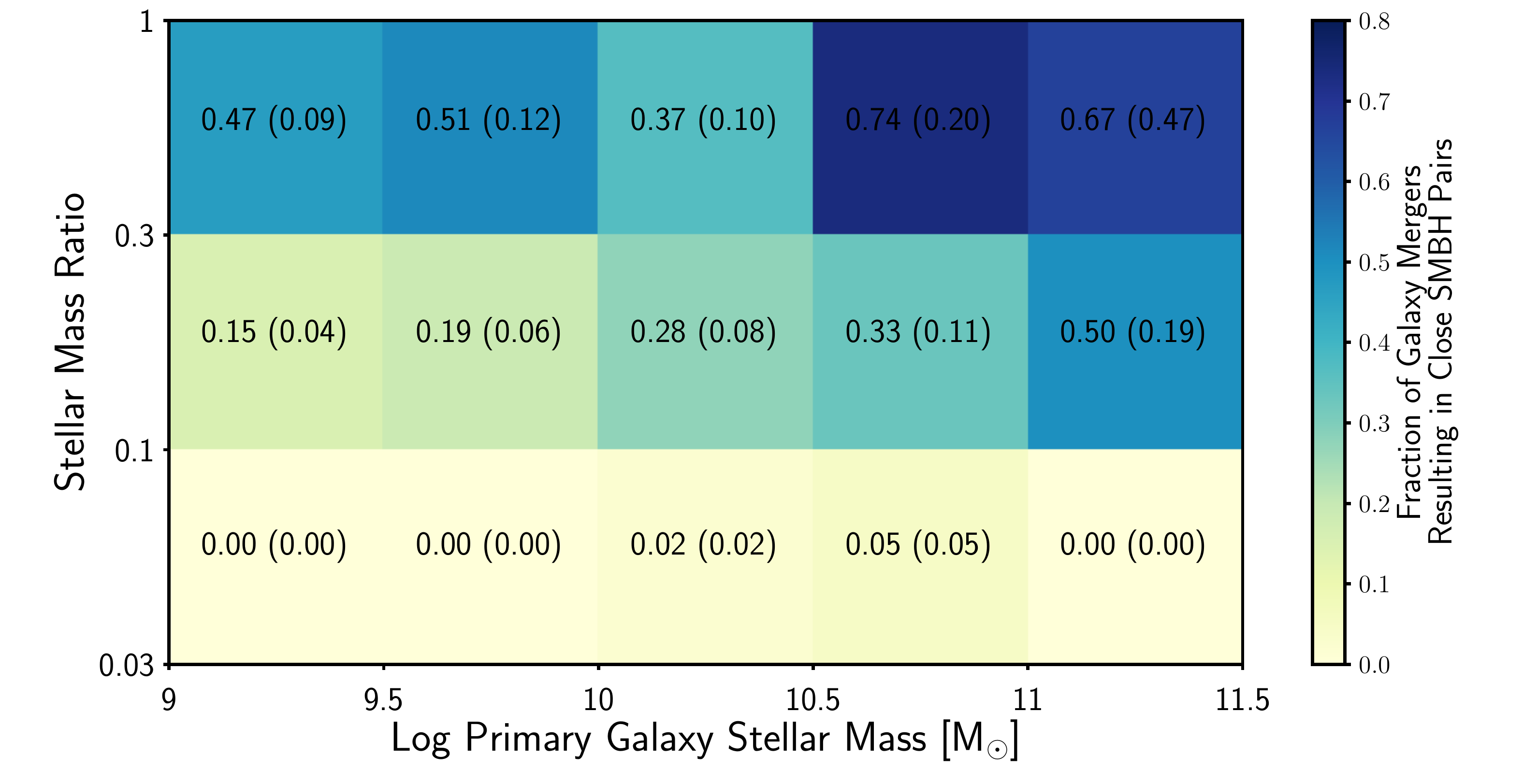}
\caption{{\sc Likelihood of Close SMBH Pair Formation}. The fraction of all galaxy mergers that result in a close SMBH pair as a function of the stellar mass of the primary galaxy and the stellar mass ratio at the time of first satellite in-fall. In addition to the colours, the fraction and, in parenthesis, associated uncertainty ($n_{pairs,i}^{0.5}/n_{i}$, where $n_i$ is the total number of galaxy mergers in each bin and $n_{pairs,i}$ the number of close SMBH pairs resulting from mergers in each bin) are labeled. Considered are galaxy mergers resulting from initial satellite in-fall at $z<5$. The formation of a close SMBH pair is not a common result of galaxy mergers. The likelihood of a close SMBH pair forming is sensitive to both stellar mass and mass ratio, and most likely to occur in massive, major mergers.}
\label{binfrac}
\end{figure}

This is a critical improvement over standard approaches to correcting SMBH dynamics that involve repositioning or pushing SMBHs toward their local potential minima \citep[e.g.][]{dimatteoBH05,diMatteo2008,genel14,eagle15}. Such methods force an un-physically fast sinking timescale for accreted SMBHs, leading to a nearly immediate formation of a close SMBH pair that does not sample the properties or kinematics of the merging galaxies \citep{tremmel15}. With this technique, the dynamics and morphology of the merging galaxies are self-consistently accounted for in the SMBH sinking timescales and the subsequent formation (or not) of a close SMBH pair.

 \subsection{Formation of Close SMBH Pairs}
 
SMBHs are assumed to form a close pair when they become closer than two softening lengths ($\approx700$ pc in our simulations) with relative velocities small enough such that they can be considered bound, i.e. $\frac{1}{2}\Delta \textbf{v} < \Delta \textbf{a} \cdot \Delta \textbf{r}$, where $\Delta \textbf{v}$, $\Delta \textbf{a}$, and $\Delta \textbf{r}$ are the relative velocity, acceleration, and distance vectors between two SMBH particles. Below this distance limit, the simulation fails to resolve the relevant stellar and gas dynamical processes involved in SMBH pair evolution and such calculations are not attempted.

In the simulation, once two SMBHs form a close pair, they are taken to act as a single SMBH with the sum of the masses. While there are still many theoretical uncertainties in the timescales to form and merge a binary SMBH system, binary hardening timescales can be relatively quick, on the order of $10^7-10^8$ yrs, if even a small amount of gas is present \citep{armitage02,haiman09,colpi14}, and even in some cases for gas poor systems \citep{holleybockelmann15}. If the binary hardening timescales are significant compared to the relevant timescales of the simulation, because the smallest resolved scales are much larger than the typical binary separation, taking the pair to act as a single object with respect to accretion and feedback is still a reasonable approximation for those processes. The timescales that we predict in the following sections are therefore a lower limit to the timescales to form a SMBH binary and subsequent merger.

We predict that the formation of close SMBH pairs is a relatively rare occurrence, with an average formation rate per co-moving volume of $0.013$ cMpc$^{-3}$ Gyr$^{-1}$. Figure~\ref{binfrac} shows the likelihood that the merger of two galaxies will result in the formation of a close SMBH pair within a Hubble time. With our formation scheme (\S2.2), lower mass galaxies are naturally less likely to host SMBHs and so often their mergers do not result in any close pairs, as one more of the galaxies do not host any SMBHs to begin with. In addition, as we explore in the next section, galaxies in lower mass ratio mergers are more likely to become tidally disrupted and deposit their SMBHs on very wide orbits with larger sinking timescales. While we will focus in the following sections on close SMBH pairs that do form in the simulation,  it is important to note that only a fraction of galaxy mergers result in a close SMBH pair forming within a Hubble time.

%
  
 \section{Close SMBH Pair Formation Timescales}
 
While several different timescales are important for understanding the formation and evolution of SMBH pairs, the evolution of SMBH orbits on kpc scales is often simplified, relying on analytic approximations that do not self consistently account for the kinematics and internal properties of the merging galaxies \citep[e.g.][]{dvorkin17}, which previous studies have shown can have an important role in determining how the SMBHs will evolve following a galaxy merger \citep[.e.g][]{G94,callegariBH09,callegari11}. With the realistic model of SMBH dynamics included in {\sc Romulus25}, the simulation is uniquely capable of estimating this timescale for a realistic population of galaxy mergers taking place within a fully cosmological environment.

For our analysis we measure the time that each eventual close pair of SMBHs spends at `galaxy-scale' ($\sim1-10$ kpc) separations. Position information for each SMBH is recorded every 1.6 Myr and simulation snapshots are recorded every $10-100$ Myr, with higher time resolution at earlier epochs. In our analysis we only include close SMBH pairs formed within resolved DM halos, with at least 10,000 DM particles, resulting in a lower mass limit of $\sim3\times10^9$ M$_{\odot}$.  We also only include close pairs that form at least 100 Myr after each SMBH has been seeded, in order to avoid counting pairings that occur as a result of multiple SMBHs forming from the same cloud of gas, a rare but possible result of our formation scheme and should be considered degenerate to a single SMBH growing quickly from a particularly large, dense cloud of gas. We confirm that our results are insensitive to the specific choice of this time threshold.

  \begin{figure}
\centering
\includegraphics[trim=0mm 0mm 10mm 0mm, clip, width=85mm]{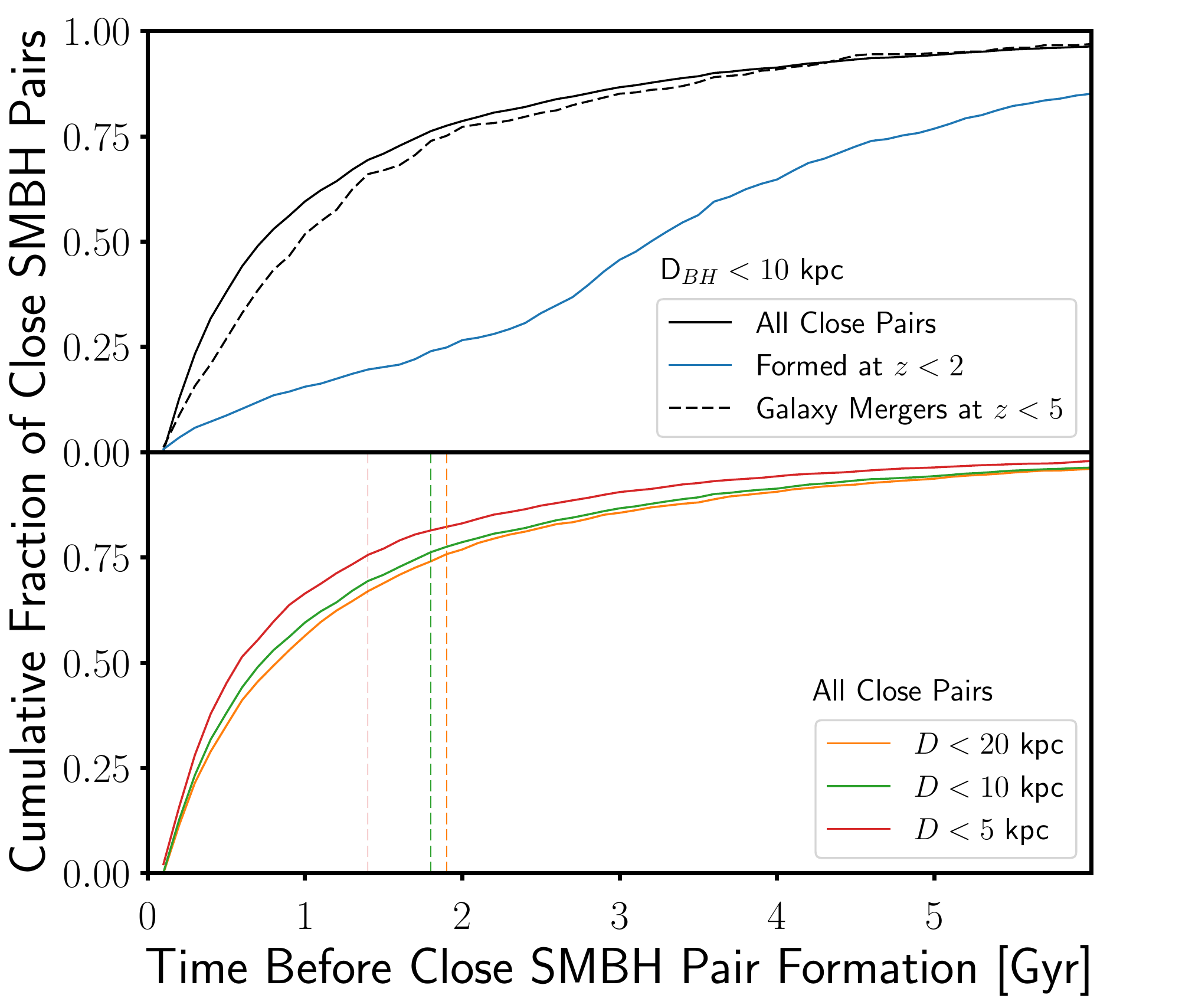}
\caption{{\sc The Timescale to Form Close SMBH Pairs.} \textit{Top:} The cumulative distribution of time that SMBH pairs spend separated by less than 10 kpc prior to close pair formation for all close SMBH pairs formed in {\sc Romulus25} (dark/black solid) While about half of the close pairs form relatively quickly ($< 0.$5 Gyr) there is a significant fraction that spend several Gyr at galaxy-scale separations. Close pairs that form at low redshift (light/blue, solid) are mostly very far removed from their progenitor galaxy merger event. Also shown is the subset of close SMBH pairs resulting satellites in-falling after $z=5$ (dashed), used in much of our analysis and which, as shown here, have timescales representative of the whole population of close SMBH pairs. \textit{Bottom:} The cumulative distribution of timescales that SMBH pairs spent at 5, (red), 10 (green), and 20 (orange) kpc separations before forming a close pair with sub-kpc separation. As expected, closer proximity implies faster sinking timescales, as the dynamical time of the galaxy at smaller radii decreases. Overall, the distributions are quite similar, implying that our results are insensitive to the specific choice of separation scales explored. Vertical dashed lines show the 75th percentiles.}

\label{tdelay_all}
\end{figure}

The top panel of Figure~\ref{tdelay_all} shows the cumulative distribution of time that SMBH pairs spend within 10 kpc of one another before forming a close pair. The distance is small enough that the two target SMBHs must be within the same galaxy or interacting pair of galaxies. For the overall population (black line) most of the close pairs form with less than 1 Gyr spent at these intermediate separations, consistent with many studies of isolated galaxy mergers \citep[e.g.][]{mayer07}. However, there is a significant population of pairs that remain at galactic-scale separations for several Gyr. Taking only the population of close pairs that form at low redshift ($z < 2$; blue line) we see that the majority of these close pairs form several Gyr after their original galaxy merger event. We therefore predict that a significant fraction of low redshift SMBH pairs (and therefore subsequent SMBH binaries and SMBH merger events) are formed from a population of long-lived, `wandering' SMBHs \citep{schneider02,volonteriOffCenBH05,bellovaryBH10} born out of early galaxy mergers.

This result can have critical implications for gravitational wave analysis in the future, affecting how such signal is interpreted in terms of connecting SMBH mergers to galaxy evolution. It can also be important for interpreting dual and offset AGN observations, as it becomes unclear how connected they may be to actual galaxy mergers. Though beyond the scope of this paper, we will explore in more detail the implications of these results to gravitational wave predictions as well as the population of offset and dual AGN in future work.
 
The bottom panel of Figure~\ref{tdelay_all}  shows the cumulative distribution of timescales that SMBH pairs spend at 5, 10, and 20 kpc separations. As expected, the evolution of SMBH pairs occurs on slightly shorter timescales for smaller separations. The sinking timescale due to dynamical friction depends on the local dynamical time, which decreases toward galactic centre. Still, we find SMBH pairs that spend several Gyrs separated by 5 kpc or less. This shows that our results are insensitive to our specific choice of separation threshold. In the following sections, we choose 10 kpc as our galaxy-scale separation threshold, as it corresponds to the size of the Galactic disk and is a good representation of the inner region of a dark matter halo that is dominated by baryonic processes. Additionally, we have confirmed that our other conclusions are also insensitive to this chosen scale.

 \begin{figure*}
\centering
\includegraphics[trim=00mm 0mm 0mm 0mm, clip, width=180mm]{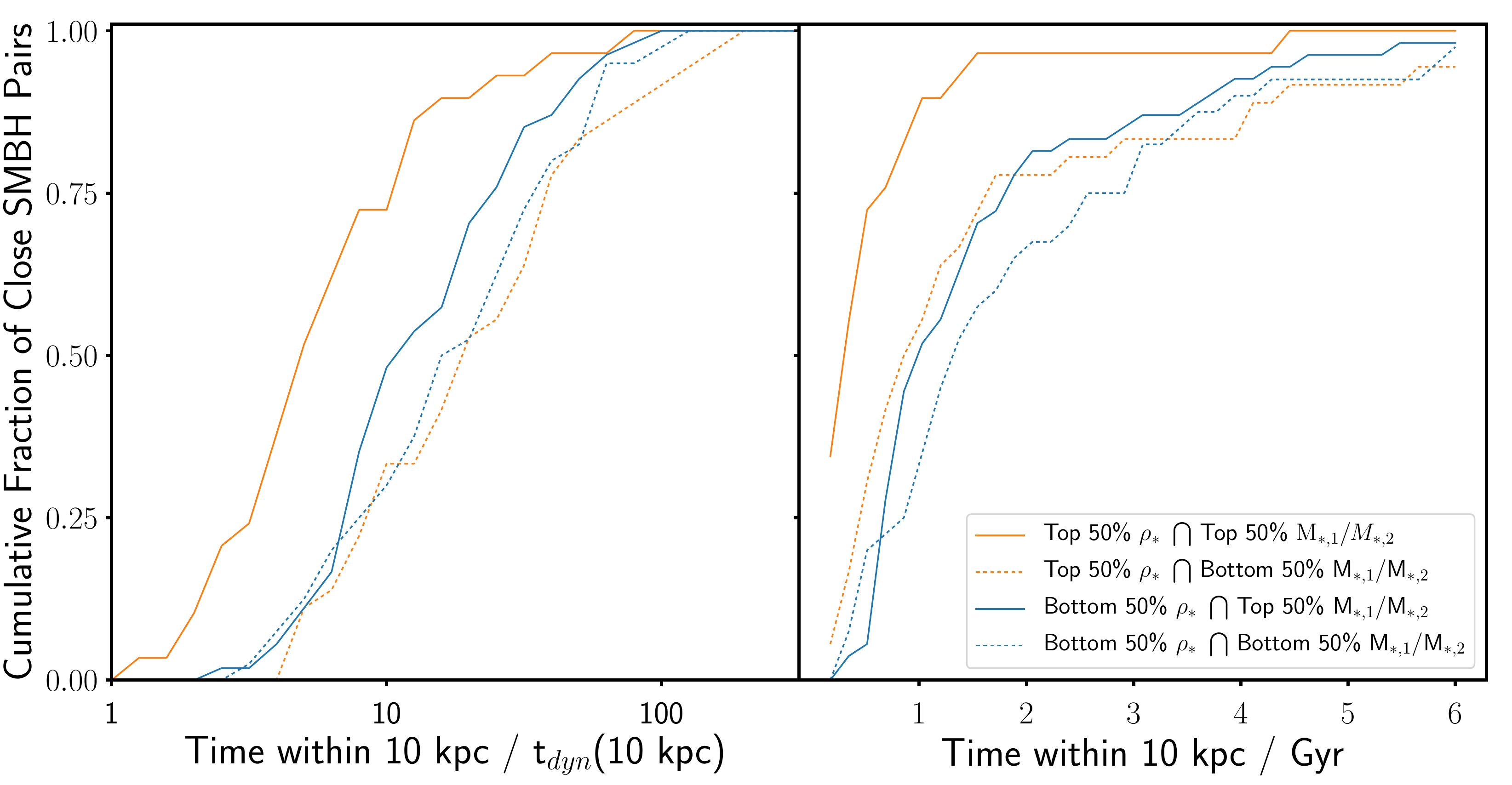}
\caption{{\sc Close Pair Formation Timescales and Merging Galaxy Properties.} The cumulative distribution of the number of dynamical times (left) and total time (right) that SMBH pairs spend within 10 kpc of one another before forming a close pair. The data is taken from 196 unique galaxy mergers taking place at $z<5$, resulting in 330 close SMBH pairs. Shown here are only those close pairs where the accreted SMBH is initially within the central 1 kpc of its host satellite galaxy (159 total pairs). The distributions are split up based on the 50th percentiles in central stellar density of the accreted galaxy and the stellar mass ratio ($3.4 \times 10^6$ M$_{\odot}$ kpc$^{-3}$ and 0.43 respectively) calculated at the in-fall time of the satellite halo. Accreted galaxies that have both high central stellar densities and high stellar mass ratios compared to the main galaxy are significantly more likely to result in a quick formation of a close SMBH pair.}
\label{tdelay_gal}
\end{figure*}

The distribution of timescales presented in Figure~\ref{tdelay_all}  is likely due to several variables, including the kinematics of the merging galaxies, the morphology of the galaxy merger remnant, the mass of the SMBHs, and where within that galaxy the SMBHs are deposited. \citet{callegari11} find that the behaviour of in-falling satellite galaxies and their host SMBHs depend strongly on the angle of the interaction. How SMBHs are deposited within a galactic disk can also affect the efficiency of dynamical friction. If the host galaxy has a cored density profile, delay timescales can also be made longer \citep{reed06,dicintio17}. Similarly, a large stellar core with high velocity dispersion could also make dynamical friction less effective, as there would be more stars moving too fast to contribute. All of these merger and galaxy properties are a natural consequence of the simulation volume and are folded into the timescale distributions we predict.

Because these timescales are the result of many different variables interacting with one anther, we find little overall dependence on single parameters like SMBH mass or halo mass. However, we do find a strong dependence on the morphology of the accreted galaxy and its stellar mass relative to the primary galaxy, which we explore in the following section.

\subsection{Galaxy Disruption and Close SMBH Pair Formation Timescales}

\begin{figure*}
\centering
\includegraphics[trim=30mm 40mm 30mm 30mm, clip, width=150mm]{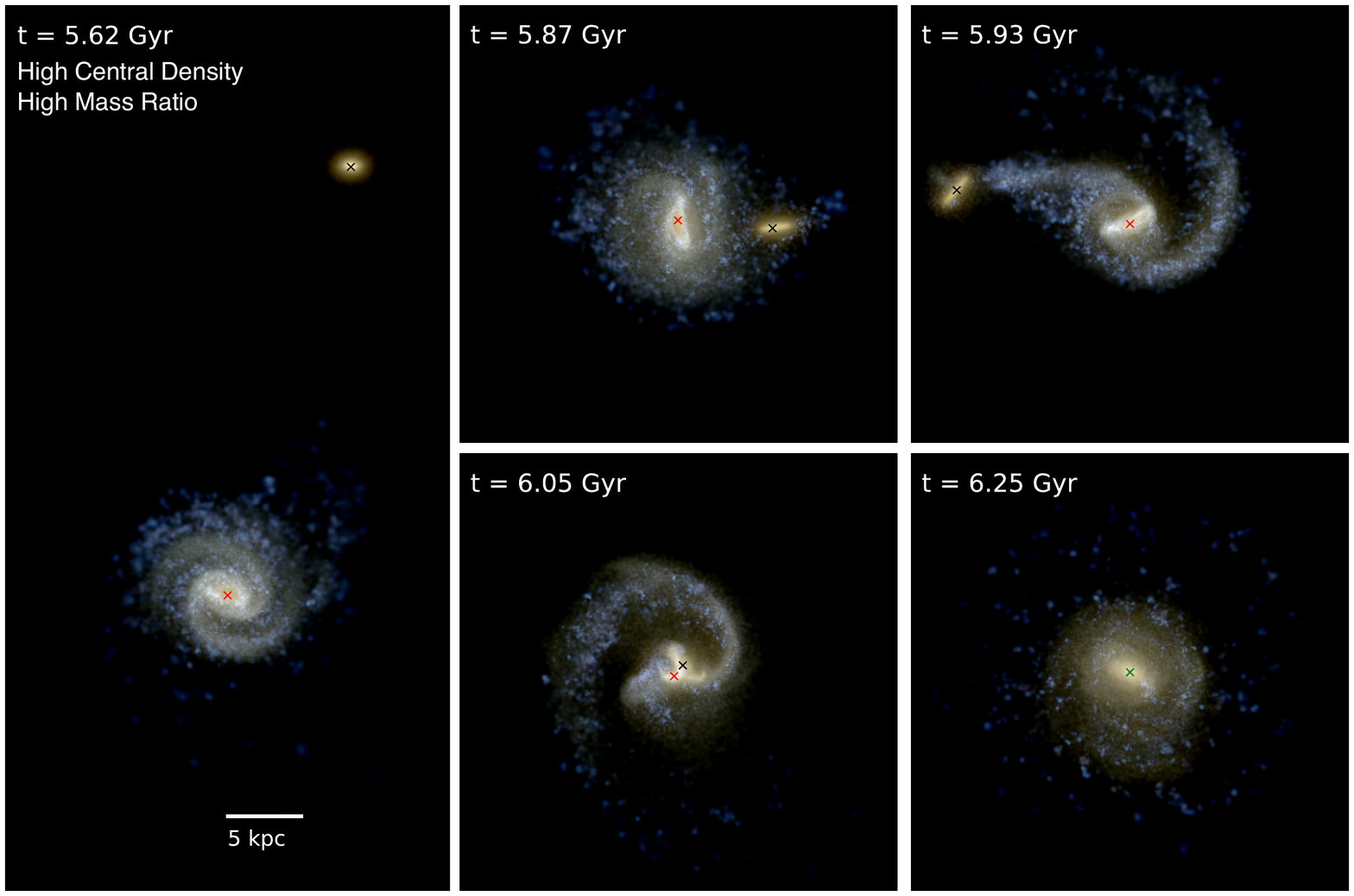}
\includegraphics[trim=30mm 40mm 30mm 30mm, clip, width=150mm]{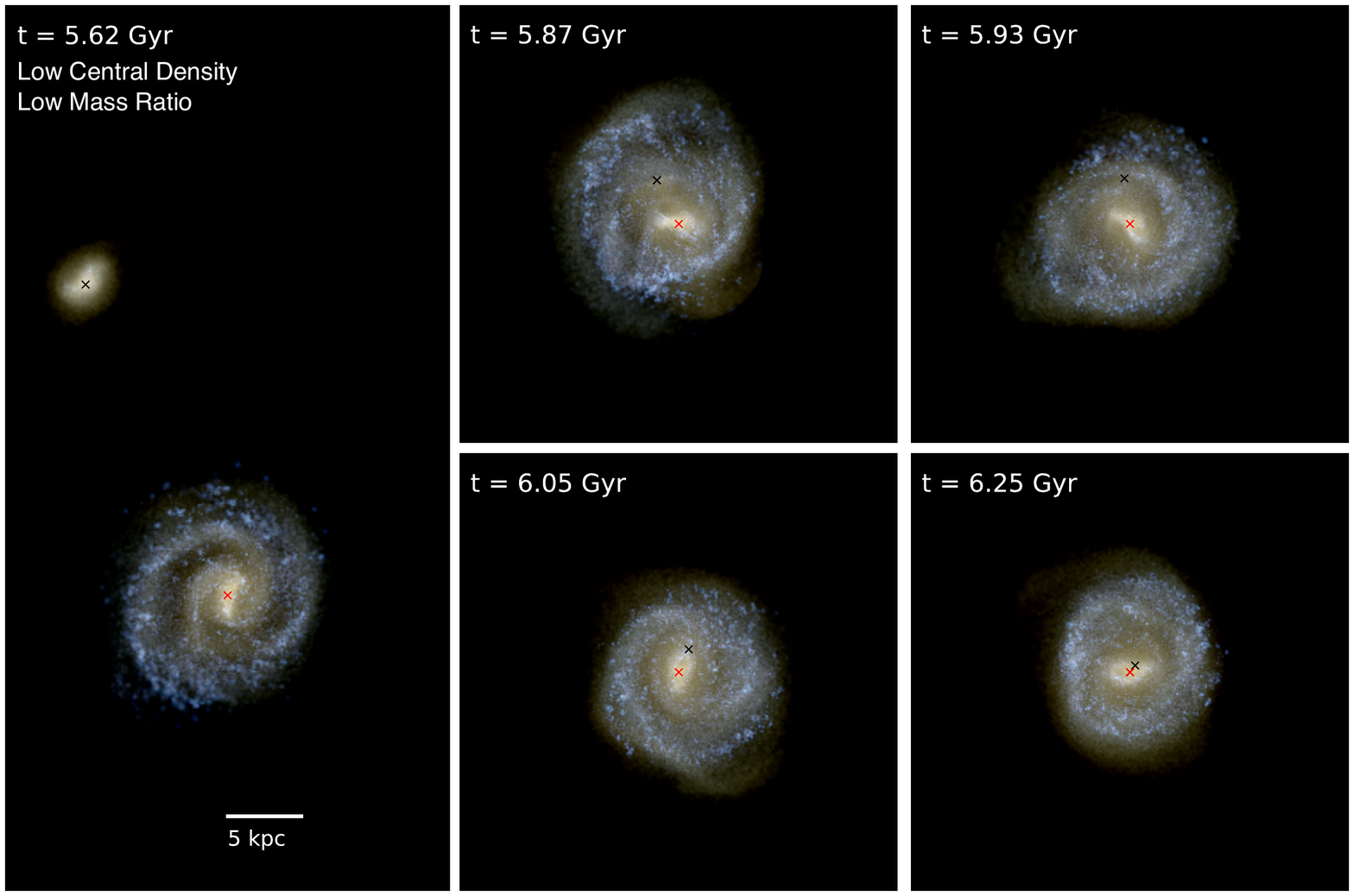}
\caption{{\sc An Illustrative Example.} Two examples of galaxy mergers taking place around the same time and with galaxies of similar mass. Each set of plots shows the spatial distribution and colour of stars at five different times leading up to and following the merger of the two galaxies. Colours are based on the contribution of different bands within each pixel using U (blue), V (green), J (red) assuming a Kroupa IMF, so young stars look blue and older stars look yellow. The stellar emission is calculated using tables generated from population synthesis models \citep[http://stev.oapd.inaf.it/cgi-bin/cmd;][]{marigo08,girardi10}. Red and black crosses mark the positions of the SMBHs and the green cross in the top final frame represents a close pair of SMBHs. The initial stellar masses of the accreted galaxies in the top and bottom cases are $1.3\times10^{10}$ and $1.02 \times 10^{10}$ M$_{\odot}$ respectively and, for the main galaxies, stellar masses of $2.9\times10^{10}$ and $4.6\times10^{10}$ M$_{\odot}$ respectively. The accreted galaxy in the top case originally has a stellar core nearly five times denser than that of the bottom galaxy. This, combined with the higher stellar mass ratio, allows the core of the galaxy to avoid disruption, quickly resulting in a close SMBH pair. In the bottom case, the core of the original galaxy is tidally heated, becomes more diffuse, and is quickly assimilated into the main galaxy, leaving the SMBH to sink on its own. Despite the close passage shown in the last frame, the SMBHs will not form a close pair until $t = 7.34$ Gyr, after $1.7$ Gyr at galaxy-scale separations compared with only $0.3$ Gyr in the top example.}
\label{examples}
\end{figure*}
 
In this section, we examine how the close SMBH pair formation timescale depends on the properties of the interacting galaxies. We take a sub-set of our close SMBH pair population that result from galaxy mergers initiated by in-falling satellites at $z < 5$, where both halos are resolved ($M_{vir}>3\times10^{9}$ M$_{\odot}$) at the time of satellite in-fall. This time of satellite in-fall is taken as the time the secondary galaxy's host dark matter halo crosses the virial radius of the main halo. For halos that cross the virial radius multiple times, the final crossing time is used. The initial properties of each galaxy prior to the merger are taken at this final in-fall time. We do not include mergers at higher redshift, as often the details of these interactions are not fully captured by our snapshots, with halos attaining a mass that passes our strict definition of what is resolved and falling into the main halo in between snapshots. This sub-sample consists of 330 close SMBH pairs resulting from 196 unique galaxy mergers. Note that, because individual galaxies can host multiple SMBHs, it is common for single galaxy mergers to result in multiple close SMBH pairs. The dashed black line in Figure~\ref{tdelay_all} shows the distribution of delay timescales for this subset of close SMBH pairs, showing that this population is indeed representative of the whole.
 
In Figure~\ref{tdelay_gal} we plot the cumulative distribution of time that eventual close SMBH pairs spend within 10 kpc of one another. We group these pairs based on the central stellar density of the in-falling galaxy and the stellar mass ratio of the two merging galaxies. The stellar density is calculated within the central kpc of each in-falling satellite galaxy. Figure~\ref{tdelay_gal} shows the results in units of both Gyr (right) and number of dynamical times (left), where the dynamical time is calculated at a radius of 10 kpc of the main galaxy at the approximate time the two SMBHs come within 10 kpc of one another. The median values for the central stellar density and stellar mass ratio are $3.4\times10^6$ M$_{\odot}$ kpc$^{-3}$ and 0.43 respectively. Only systems where the accreted SMBH is within the central 1 kpc of its host galaxy at in-fall time are considered. Initially offset SMBHs are considered in the next section.

It is clear from this figure that accreted galaxies with high central densities and higher stellar mass ratios result in significantly shorter delay times. Galaxies with either low central densities or low stellar mass ratios experience longer times spent at galaxy-scale separations, implying that tidal disruption of the host galaxy is important for determining the timescale for close SMBH pair formation. During a galaxy interaction, ram pressure stripping can disrupt gas within galactic disks at larger radii and tidal heating can disrupt the inner core of the galaxies. Dense stellar cores within high mass ratio mergers are more likely to avoid disruption through both ram pressure stripping and tidal heating \citep{gnedin99,callegariBH09,vanWassenhove14}, so the central SMBHs remain embedded in a dense stellar core that aids in their orbital decay.  In galaxies lacking a dense stellar core, or those involved in more minor mergers, tidal heating is more efficient at disrupting the inner parts of the galaxy, resulting in SMBHs deposited at large radii without any stellar core to assist in their orbital decay. This is consistent with analytical experiments showing how the orbital evolution of SMBHs is highly dependent on whether they are embedded in a stellar core or `naked' within their new host galaxy \citep{Yu02, dosopoulou17}.

Figure~\ref{examples} shows a series of snapshots from two example galaxy mergers taking place with both primary and secondary galaxies initially within a factor of 2 of one another in stellar mass. However, the stellar mass ratio of the top and bottom examples is 0.45 and 0.22 respectively. This, combined with the fact that the secondary galaxy in the top example has an initial central stellar density nearly 5 times higher than that in the bottom case, results in very different SMBH orbital evolution. In the bottom case,  the secondary galaxy's core becomes tidally heated and eventually disrupted by the main galaxy, no longer maintaining its structure. In the top case, the denser core is able to avoid disruption and maintains its integrity up until the two cores merge, bringing the SMBHs along with them. The bottom example of a disrupted galaxy forms a close SMBH pair only after the SMBHs spend $1.7$ Gyr within 10 kpc of one another, while the top case results in a close pair after the SMBHs spend only 0.3 Gyr at galaxy-scale separations.

\subsection{Initially Offset SMBHs}

In the previous section, we focused on central SMBHs, those that are at the centre of their host galaxy at the time of satellite in-fall. However, approximately half of the close SMBH pairs in our sub-sample from {\sc Romulus25} form from accreted SMBHs initially offset from the centre of their host galaxy. As we have seen, the orbital decay of SMBHs can often take several Gyr and galaxy mergers often never result in a close SMBH pair. Massive galaxies in the {\sc Romulus25} simulation therefore often have several SMBHs that are offset from galactic centre, gathered throughout the host galaxy's merger history. In some rare cases, galaxies only have offset SMBHs. 

\begin{figure}
\centering
\includegraphics[trim=0mm 0mm 0mm 0mm, clip, width=85mm]{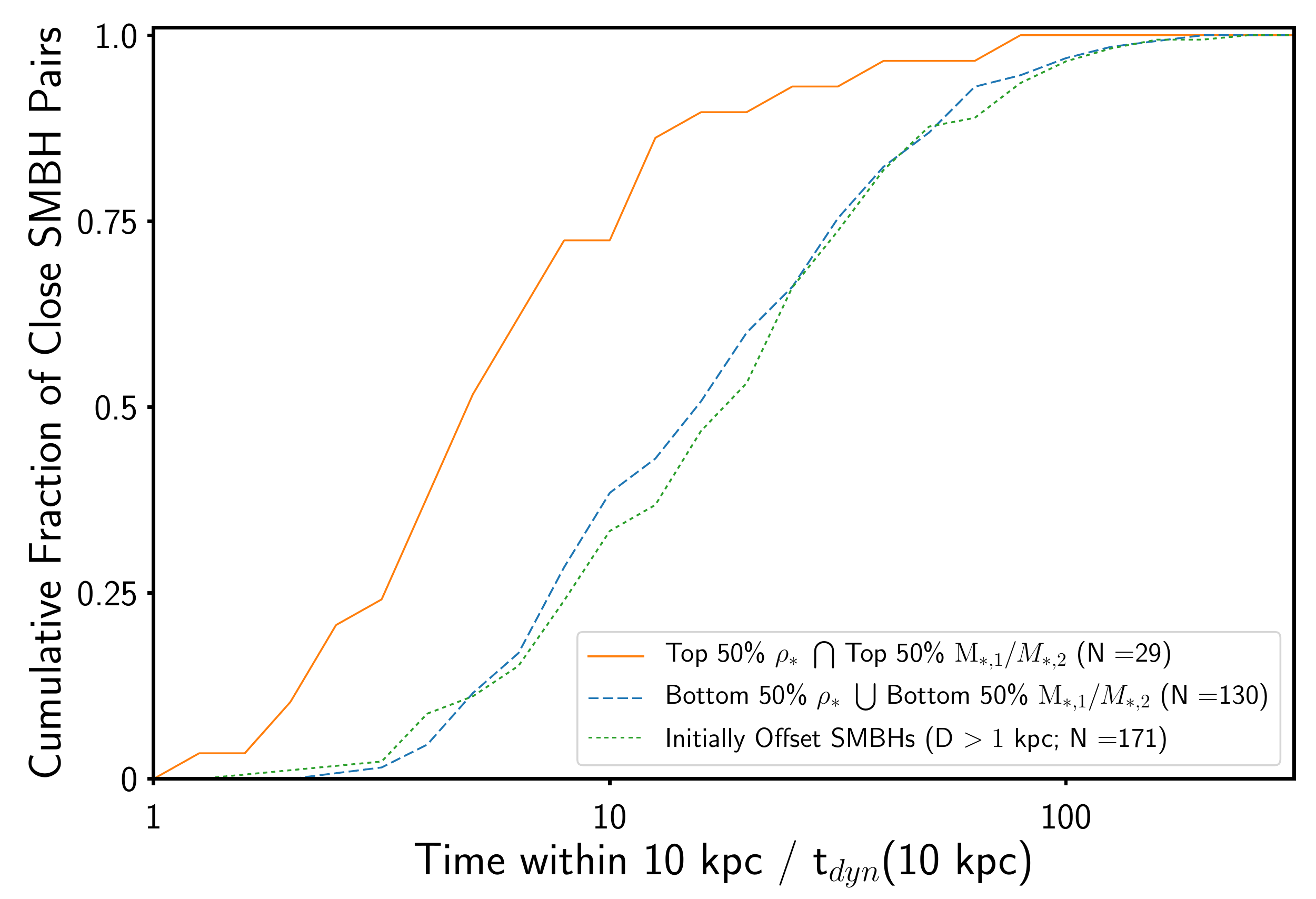}
\caption{{\sc Close Pair Formation Timescales for Initially Offset SMBHs.} The cumulative distribution of the number of dynamical times SMBH pairs spend within 10 kpc of one another before forming a close pair. The solid orange line represents SMBHs from galaxies that are less susceptible to disruption (same as in Figure~\ref{tdelay_gal}) and the blue dashed line represents SMBHs from galaxies that are more likely to become tidally disrupted due to a lower stellar mass ratio and/or low central density (the union of the other three lines shown in Figure~\ref{tdelay_gal}). The green dotted line represents SMBHs that were initially offset from the centres of their host satellite galaxies by more than 1 kpc at the time of in-fall. The green and blue distributions are very similar, which is to be expected. In both cases, the SMBHs lack the extra support of a stellar core when making their way to the centre of their new galaxy.} 
\label{tdelay_off}
\end{figure}

Figure~\ref{tdelay_off} is similar to the left panel of Figure~\ref{tdelay_gal}, with SMBH binaries binned based on whether the accreted host galaxy is more likely to avoid complete disruption due to a dense stellar core and high mass ratio (orange/solid), less likely to avoid disruption (blue/dashed), or whether the target SMBH is offset from the centre of their host satellite galaxy by more than 1 kpc as it crosses the main halo's virial radius (green/dotted).

The close pair formation timescale distribution for initially offset SMBHs is similar to that for more easily disrupted satellite galaxies. When the SMBH is not central, it is likely not embedded within a dense stellar core, even if its host galaxy has one. It will therefore become accreted onto the main galaxy without a stellar core to aid in dynamical friction, just like SMBHs in galaxies whose cores become tidally disrupted.

\subsection{The Importance of Galaxy-Scale Orbital Evolution}

The previous sections have shown that SMBH pairs can spend significant time at kpc-scale distances before forming a close pair.  Their evolution on kpc scales is a phase that is very difficult to fully capture analytically, as it straddles the separations where galaxies are still merging and those where the sinking concerns the SMBHs themselves, naked or surrounded by the core of their satellite (see a discussion in \citet{mcwilliams14}). When estimating the time of binary formation (for which the time of pair formation studied here is a lower limit) semi-analytical models normally use satellite merging timescales that should account for the full `amalgamation' of the satellite (see a discussion in \citet{bk08}), typically estimated from large suites of dark matter-only simulations. \citet{bk08} argue that the inclusion of baryons (specifically, bulges, that are denser than dark matter and thus more resistant to disruption) would shorten the timescales compared to estimates for dark matter haloes alone. The results of the previous sections, however, show a more complex picture when dealing with SMBHs, rather than halos and galaxies only. 

In order to test the approach of semi-analytic models, we estimate the close pair formation times that would be predicted from more simplistic models. We approximate the halo sinking timescale via the analytic fit derived by \citet{bk08}, given the halo masses of the primary and satellite halos and the virial radius of the primary halo taken from the simulation at the time of satellite in-fall. Following a procedure similar to modern semi-analytic models \citep[e.g.][]{Barausse12}, we give each halo pair a circularity, $\epsilon = j / j_{circ}$, sampled from a normal distribution centred at $\bar{\epsilon} = 0.5$ and with $\sigma = 0.23$ \citep{Khochfar06}.  The circular radius is calculated from the periastron radius, approximated by $r_{peri} = R_{vir} \epsilon^{2.17}$ \citep{Khochfar06}. In order to remain in the regime where the fit from \citet{bk08} is accurate, we only allow $\epsilon$ to vary between 0.2 and 1.0. Below $\epsilon=0.2$, baryonic effects dominate due to the satellite galaxy's very radial orbit, making the approximation less accurate. This simple approach allows us to compare the sinking times predicted from {\sc Romulus25} to the average halo sinking timescales that would be included in most semi-analytic models.

\begin{figure}
\centering
\includegraphics[trim=10mm 0mm 20mm 20mm, clip, width=85mm]{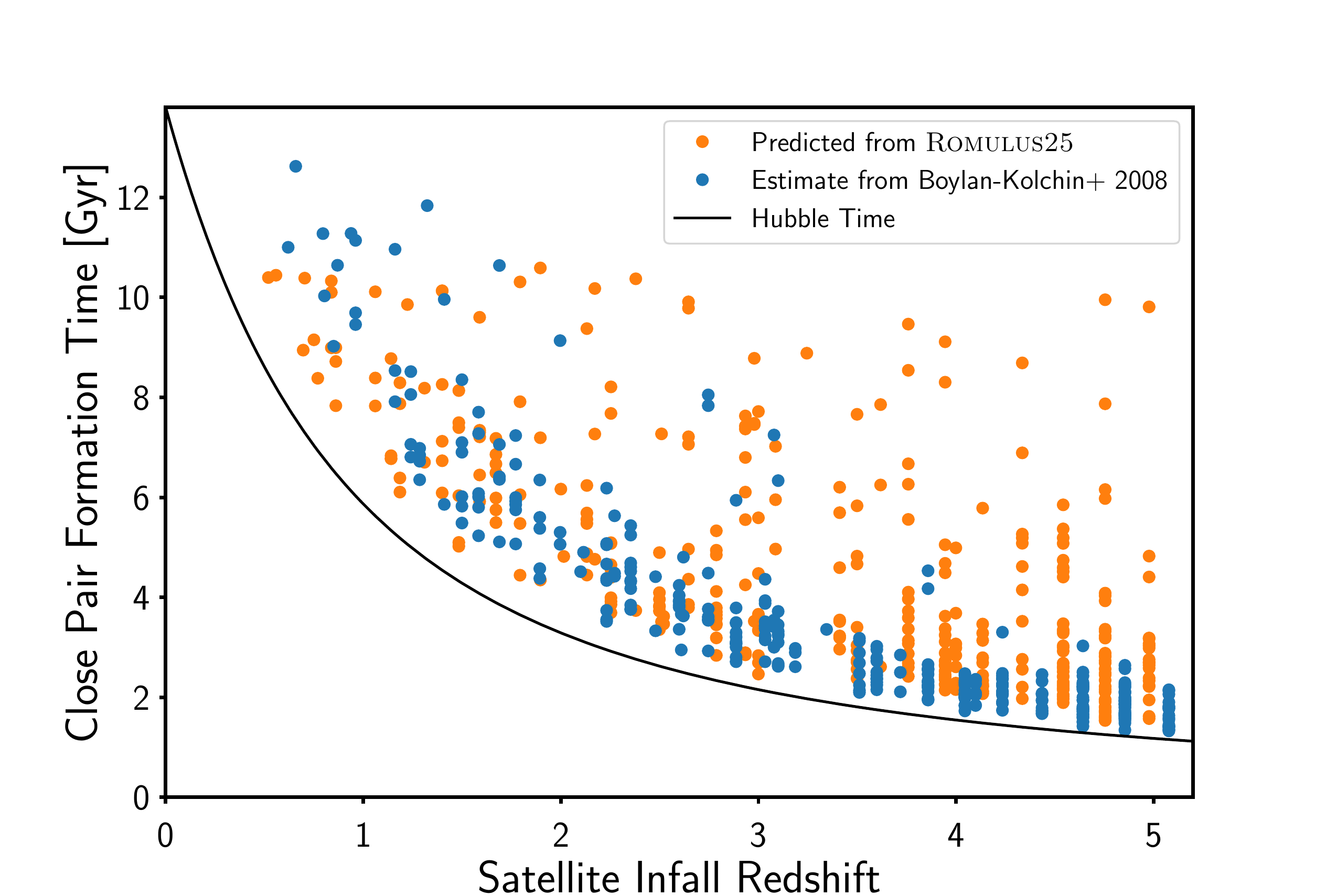}
\caption{{\sc SMBH vs. Halo Sinking Timescales.} The formation time of close SMBH pairs as a function of satellite in-fall redshift. The black line denotes the time as a function of redshift. The orange points plot the time of close SMBH pair formation predicted directly from the {\sc Romulus25} simulation. The blue points estimate what the close pair formation time would be only accounting for halo sinking timescales approximated by the analytic fit from \citet{bk08}, as described in the text. The in-fall redshifts are shifted slightly between the two in order to make the distinction more clear. For high redshift halo mergers, the timescale for SMBH orbits to decay from galaxy-scale separations is a critical bottleneck to close pair formation, resulting in formation times that are often much later than those predicted solely based on halo sinking timescales. At lower redshift ($z<2$) the halo sinking timescales represent an increasingly important bottleneck to the formation of close SMBH pairs, resulting in less difference between the two types of points.}

\label{tdelay_comp}
\end{figure}

We find that galaxy-scale orbital evolution is an important bottleneck to close SMBH pair formation for high redshift galaxy mergers. In Figure~\ref{tdelay_comp} we plot the close pair formation times directly from the {\sc Romulus25} cosmological simulation against the in-fall redshift of the parent satellite galaxy for the secondary SMBH (orange points). We compare this time to that which would be predicted solely using the analytic halo sinking timescale described above (blue points). In other words, these points represent the time for close pair formation if galaxy-scale orbital evolution and other baryonic effects were ignored, as they often are in both semi-analytic models and other cosmological simulations.  We find that the orbital evolution of SMBHs from 10 kpc to sub-kpc scales is an important bottleneck to close pair formation (and the subsequent binary formation and merger) for high redshift galaxy interactions, where the dynamical timescale for satellite halos is comparatively small. At  redshift less than $\sim2$ we find that there is less of a clear difference between the two types of points, indicating that halo sinking timescales are more similar to or even sometimes dominant compared to galaxy-scale SMBH orbital evolution. Semi-analytic models of SMBH binary evolution find similar results, with binary evolution timescales acting as a dominant bottleneck at high redshift and increasingly less important when compared to satellite sinking timescales at low redshift \citep{volonteri16IAUb}.


Examining the halo in-fall times and the predicted close SMBH pair formation it is clear that close pairs that form at later times are often the consequence of high redshift mergers, an effect also seen in Figure~\ref{tdelay_all}. These results show that SMBH orbital evolution on galaxy scales is a very important bottleneck for the formation of close SMBH pairs and, therefore, SMBH binaries and mergers, and must be accounted for when predicting the population of binary SMBHs and gravitational wave events across cosmic time.



\section{Discussion and Conclusions}

Using the {\sc Romulus25} cosmological simulation, which is uniquely capable of tracking the dynamics of SMBHs within galaxies down to sub-kpc scales, we examine the timescale for SMBH pairs to evolve from galaxy-scale separations ($1-10$ kpc) to form close pairs with separations less than a kpc, the precursor phase to a bound SMBH binary and (possible) future SMBH merger. The formation of close SMBH pairs is a relatively rare occurrence, becoming more common in major mergers of more massive galaxies.  We find that galaxy mergers across cosmic time result in close SMBH pairs that often form several Gyr after the original galaxy merger event. SMBHs often accrete onto a new host galaxy via galaxy merger at high redshift, but only form a close SMBH pair at much lower redshift, resulting in a long lived population of `wandering' SMBHs \citep{schneider02,volonteriOffCenBH05,bellovaryBH10}. This can affect how we predict and interpret future observations of gravitational waves and dual/offset AGN, as well as the observational signatures of gravitational recoil events \citep{blecha16}. 

Using a set of 330 SMBH close pairs resulting from 196 unique galaxy mergers within {\sc Romulus25}, we show that the timescales for the formation of a close SMBH pair is dependent on galaxy morphology and stellar mass ratio. Galaxy mergers with similar mass and dense stellar cores result in faster close pair formation, as the secondary galaxy is less likely to become tidally disrupted. SMBHs that are embedded in stellar cores that are able to avoid disruption will be aided in sinking to galactic centre \citep{Yu02, dosopoulou17}. Satellite galaxies that are more susceptible to tidal disruption result in longer SMBH sinking timescales and close SMBH pairs that form long after the galaxy merger event (if they form at all). A similar situation is true for SMBHs that are initially offset from the centre of satellite galaxies. These SMBHs are not likely to be within the central stellar core, if one exists, of their host galaxy and so are deposited on their own at relatively large radii during the galaxy interaction.

The resolution limit of the {\sc Romulus25} simulation affects the scale at which tidal heating and disruption can be captured. Tidal processes become important when the impact parameter is similar to the effective radius of the disrupting object. With a Plummer equivalent gravitational force resolution of 250 pc, the effective radius of galaxies are well resolved for a wide range of masses and redshifts \citep{graham08,vanderwel14} and so disruption occurring on large scales is captured, but the internal structure on scales very close to the SMBHs remains unresolved. Dense cusps of stars can form in galaxies, particularly during gas rich mergers. These dense regions would persist for longer, as they require closer interactions to tidally heat. These unresolved stellar remnants can have an important effect on SMBH dynamics on scales much lower than 700 pc \citep{vanWassenhove14}, the limit beyond which we do not attempt to follow them in this work.

SMBHs deposited on larger scales may still have a dense stellar core or nuclear star cluster \citep{wehner06,ferrarese06b} around them that {\sc Romulus25} is unable to resolve, effectively increasing their dynamical mass. However, for the sample of close SMBH pairs formed from galaxy mergers where disruption likely takes place, we find that the sinking time does not show a clear dependence on SMBH mass. This indicates that the existence of an unresolved, dense stellar component around these SMBHs will only have a secondary effect on their orbital evolution. Rather, the sinking times depend more on the details of the galaxy merger, i.e. where and with what orbital energy the SMBHs deposited.

We show that orbital evolution of SMBHs within galaxies on scales between 1-10 kpc are a major bottle neck for forming close SMBH pairs, particularly for high redshift galaxy interactions. In agreement with the arguments by \citet{volonteri16IAUb}, at lower redshifts ($z<2$) the sinking timescale of satellite halos becomes a more dominant factor and the specific effect of galaxy-scale orbital decay is less important, though still not trivial. How much of an effect this timescale plays in the overall prediction for SMBH merger rates will also depend on the hardening timescales after formation of the binary. While there is evidence that such hardening times can be relatively short, on the order of $10^7-10^8 yr$ \citep{armitage02,haiman09,colpi14,holleybockelmann15}, other recent work suggests that these hardening timescales may be very long in some cases \citep{vasiliev15,kelley17,tamburello17}. Further, it is important to note that we do not include the effects of gravitational recoil, nor three-body SMBH encounters, both of which can further affect the formation of SMBH binaries.

It is clear that this stage of SMBH pair evolution plays a crucial role in determining when and where close SMBH pairs occur, and therefore the SMBH binaries and mergers that may result from such pairs. It is also important to understanding  the time connection between AGN activity and galaxy interaction induced star formation, as the SMBH sinking timescale may be much larger than that of the typical observed starburst timescale, found to be on the order of 0.1 Gyr \citep{marcillac06,pereira-santaella15}. As illustrated in Figure~\ref{examples}, close SMBH pairs often form in relaxed galaxies that show no morphological disturbances indicative of a recent merger. 

In future work, we will examine in more detail how this additional timescale can affect SMBH merger predictions from state-of-the-art SAMs, exploring in particular how the close pair formation timescale explored in this work compares with other affects such as three body interactions and binary hardening rates in determining the predicted signals for future gravitational wave observatories. We will also explore the occurrence of dual and offset AGN \citep[e.g.][]{comerford14,comerford15,barrows17}, to examine in more detail the phase of galaxy and SMBH evolution traced by these events.

 \section*{Acknowledgments}

The Authors thank the anonymous referee for a thorough reading of the manuscript and their helpful comments. FG, TQ and MT were partially supported by NSF award AST-1514868.  AP was supported by the Royal Society. This research is part of the Blue Waters sustained-petascale computing project, which is supported by the National Science Foundation (awards OCI-0725070 and ACI-1238993) and the state of Illinois. Blue Waters is a joint effort of the University of Illinois at Urbana-Champaign and its National Center for Supercomputing Applications. This work is also part of a PRAC allocation support by the National Science Foundation (award number OCI-1144357). MV acknowledges funding from the European Research Council under the European Community's Seventh Framework Programme (FP7/2007-2013 Grant Agreement no. 614199, project `BLACK'). Much of the analysis done in this work was done using the software packages Pynbody \citep{pynbody} nd TANGOS (Pontzen et al., in prep). The authors thank Priyamvada Natarajan, Angelo Ricarte, Enrico Barausse, Laura Blecha, Julie Comerford, and Lisa Steinborn for stimulating discussions and a careful reading of the manuscript.

\bibliography{bibref_mjt.bib}
\bibliographystyle{mn2e}

\appendix

\section{Dynamical Friction Test at Romulus25 Resolution}


In this section we explicitly confirm that the dynamical friction prescription presented in \citet{tremmel15} is able to correctly track the orbital decay of a SMBH at the resolution of {\sc Romulus25}. To do this, we set up a similar experiment to that presented in \citet{tremmel15}. We run an idealized simulation of an isolated, collapsing over-density using the publicly available software, ICInG\footnote[1]{https://github.com/mtremmel/ICInG.git}. The simulation is dark matter only with a particle mass of $3.2\times10^5$ M$_{\odot}$ and gravitational softening, $\epsilon_g$, of 342 pc, within 10\% and 3\% of the values used in {\sc Romulus25} respectively. We set up and run the initial overdensity collapse until its virial mass is $3.6\times10^{11}$ M$_{\odot}$, but larger scales are still actively collapsing, as in a cosmological simulation. At this time, the density profile is consistent with an NFW profile \citep{navarro96} of concentration 6 and a virial radius of 143 kpc. In order to test pair formation timescales, we run one simulation with two SMBHs. One is a central SMBH and one is an off-center SMBH on an eccentric orbit. Both SMBHs are $10^6$ M$_{\odot}$. The off-center SMBH is placed at 2 kpc from the center of the halo with a tangential velocity of 4.7 km/s relative to the center of mass velocity of the inner 5 kpc of the halo. This is approximately $0.1v_{circ}$ for an NFW halo at this radius with the test halo's size, mass, and concentration. The SMBH placed at the center is given no relative velocity. In order to more accurately model the conditions for SMBH pair formation in {\sc Romulus25}, we ensure that both SMBHs have a timestep of $\sim10^5$ yrs, similar to the largest timesteps for SMBHs in {\sc Romulus25}. In the simulation, two SMBHs are allowed to form a close pair when they are within $2\epsilon_g$ of one another because below this scale the orbital evolution is poorly resolved. Their relative velocities must also be consistent with being mutually bound. This avoids having two SMBHs form a close pair when one is on an eccentric orbit that may bring it into close proximity of a central SMBH, as in this test scenario.

\begin{figure}
\centering
\includegraphics[trim=0mm 0mm 0mm 0mm, clip, width=85mm]{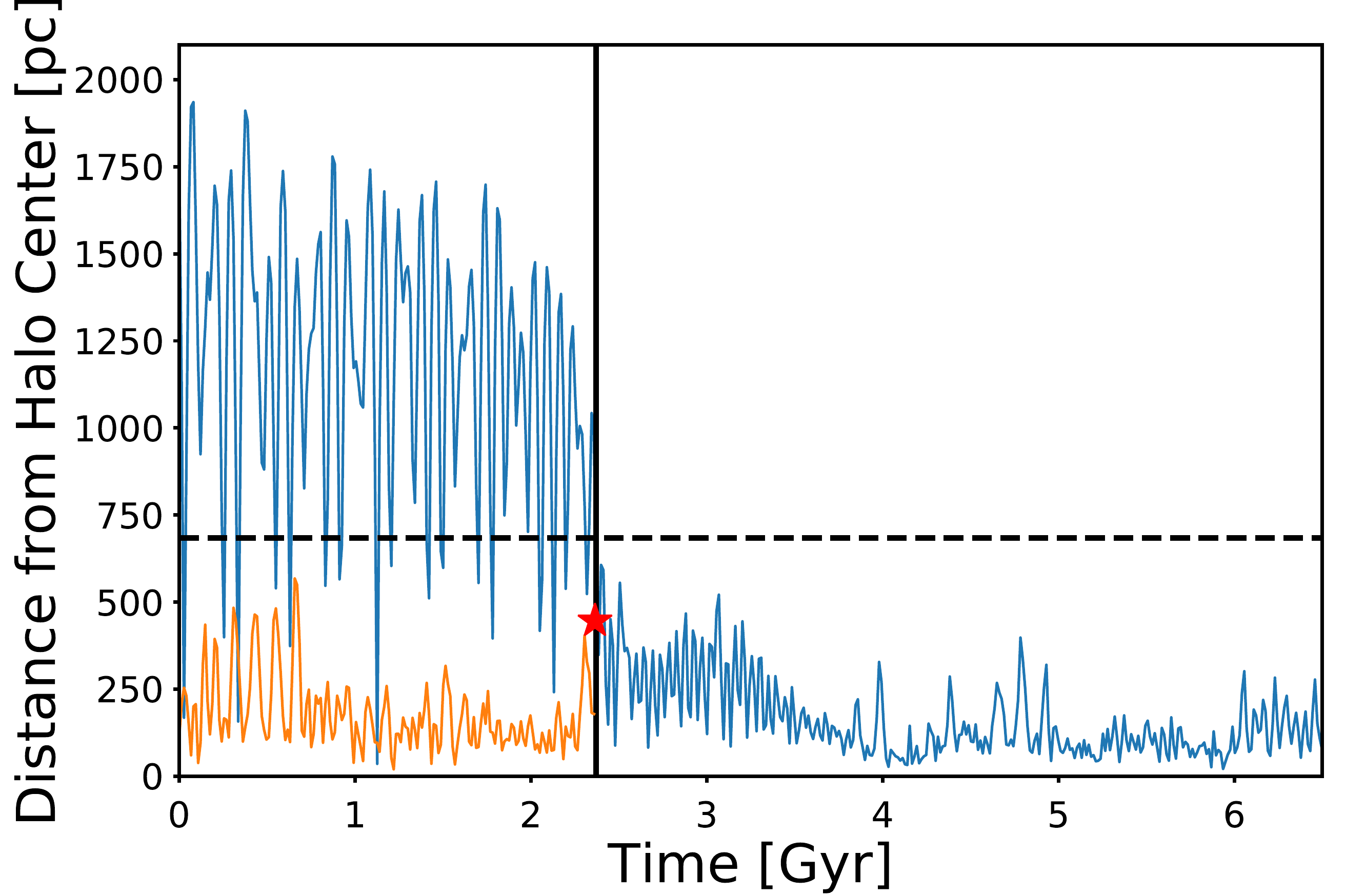}
\caption{{\sc Close Pair Formation Timescale Test.} Two SMBHs evolved within an isolated, actively collapsing dark matter halo. One SMBH is initially in the center and the other on an eccentric orbit with apocenter of 2 kpc. To ensure accurate representation of the {\sc Romulus25} simulation, the time steps for SMBHs are forced to be $\sim10^5$ yrs, similar to the largest time steps for SMBHs in {\sc Romulus25}. The dashed horizontal line represents $2\epsilon_g$ from halo center and the vertical line the theoretical dynamical friction sinking timescale, $\tau_{DF}$ from  \citet{taffoni03}. The two lines correspond to the two SMBHs and the red star the position and time when the two SMBHs form a close pair and are then tracked by a single particle with mass $2\times10^6$ M$_{\odot}$. The merger occurs at a time very nearly equal to $\tau_{DF}$.}
\label{appendix_df2}
\end{figure}

The result of this simulation is shown in Figure~\ref{appendix_df2}. The SMBH pair forms (shown as the red star) at a time \textit{nearly equal to the analytic prediction} from \citet{taffoni03} (black vertical line) with \textit{no tuning at all of our sub-grid physics}. This experiment confirms that the method used for correcting for unresolved dynamical friction, combined with our pair formation criteria, works at the resolution attained in {\sc Romulus25}. 

\label{lastpage}
\end{document}

%% file: rgb.tex
  \definecolor{snow}{rgb}{1.000000,0.980392,0.980392}
  \definecolor{ghost white}{rgb}{0.972549,0.972549,1.000000}
  \definecolor{GhostWhite}{rgb}{0.972549,0.972549,1.000000}
  \definecolor{white smoke}{rgb}{0.960784,0.960784,0.960784}
  \definecolor{WhiteSmoke}{rgb}{0.960784,0.960784,0.960784}
  \definecolor{gainsboro}{rgb}{0.862745,0.862745,0.862745}
  \definecolor{floral white}{rgb}{1.000000,0.980392,0.941176}
  \definecolor{FloralWhite}{rgb}{1.000000,0.980392,0.941176}
  \definecolor{old lace}{rgb}{0.992157,0.960784,0.901961}
  \definecolor{OldLace}{rgb}{0.992157,0.960784,0.901961}
  \definecolor{linen}{rgb}{0.980392,0.941176,0.901961}
  \definecolor{antique white}{rgb}{0.980392,0.921569,0.843137}
  \definecolor{AntiqueWhite}{rgb}{0.980392,0.921569,0.843137}
  \definecolor{papaya whip}{rgb}{1.000000,0.937255,0.835294}
  \definecolor{PapayaWhip}{rgb}{1.000000,0.937255,0.835294}
  \definecolor{blanched almond}{rgb}{1.000000,0.921569,0.803922}
  \definecolor{BlanchedAlmond}{rgb}{1.000000,0.921569,0.803922}
  \definecolor{bisque}{rgb}{1.000000,0.894118,0.768627}
  \definecolor{peach puff}{rgb}{1.000000,0.854902,0.725490}
  \definecolor{PeachPuff}{rgb}{1.000000,0.854902,0.725490}
  \definecolor{navajo white}{rgb}{1.000000,0.870588,0.678431}
  \definecolor{NavajoWhite}{rgb}{1.000000,0.870588,0.678431}
  \definecolor{moccasin}{rgb}{1.000000,0.894118,0.709804}
  \definecolor{cornsilk}{rgb}{1.000000,0.972549,0.862745}
  \definecolor{ivory}{rgb}{1.000000,1.000000,0.941176}
  \definecolor{lemon chiffon}{rgb}{1.000000,0.980392,0.803922}
  \definecolor{LemonChiffon}{rgb}{1.000000,0.980392,0.803922}
  \definecolor{seashell}{rgb}{1.000000,0.960784,0.933333}
  \definecolor{honeydew}{rgb}{0.941176,1.000000,0.941176}
  \definecolor{mint cream}{rgb}{0.960784,1.000000,0.980392}
  \definecolor{MintCream}{rgb}{0.960784,1.000000,0.980392}
  \definecolor{azure}{rgb}{0.941176,1.000000,1.000000}
  \definecolor{alice blue}{rgb}{0.941176,0.972549,1.000000}
  \definecolor{AliceBlue}{rgb}{0.941176,0.972549,1.000000}
  \definecolor{lavender}{rgb}{0.901961,0.901961,0.980392}
  \definecolor{lavender blush}{rgb}{1.000000,0.941176,0.960784}
  \definecolor{LavenderBlush}{rgb}{1.000000,0.941176,0.960784}
  \definecolor{misty rose}{rgb}{1.000000,0.894118,0.882353}
  \definecolor{MistyRose}{rgb}{1.000000,0.894118,0.882353}
  \definecolor{white}{rgb}{1.000000,1.000000,1.000000}
  \definecolor{black}{rgb}{0.000000,0.000000,0.000000}
  \definecolor{dark slate gray}{rgb}{0.184314,0.309804,0.309804}
  \definecolor{DarkSlateGray}{rgb}{0.184314,0.309804,0.309804}
  \definecolor{dark slate grey}{rgb}{0.184314,0.309804,0.309804}
  \definecolor{DarkSlateGrey}{rgb}{0.184314,0.309804,0.309804}
  \definecolor{dim gray}{rgb}{0.411765,0.411765,0.411765}
  \definecolor{DimGray}{rgb}{0.411765,0.411765,0.411765}
  \definecolor{dim grey}{rgb}{0.411765,0.411765,0.411765}
  \definecolor{DimGrey}{rgb}{0.411765,0.411765,0.411765}
  \definecolor{slate gray}{rgb}{0.439216,0.501961,0.564706}
  \definecolor{SlateGray}{rgb}{0.439216,0.501961,0.564706}
  \definecolor{slate grey}{rgb}{0.439216,0.501961,0.564706}
  \definecolor{SlateGrey}{rgb}{0.439216,0.501961,0.564706}
  \definecolor{light slate gray}{rgb}{0.466667,0.533333,0.600000}
  \definecolor{LightSlateGray}{rgb}{0.466667,0.533333,0.600000}
  \definecolor{light slate grey}{rgb}{0.466667,0.533333,0.600000}
  \definecolor{LightSlateGrey}{rgb}{0.466667,0.533333,0.600000}
  \definecolor{gray}{rgb}{0.745098,0.745098,0.745098}
  \definecolor{grey}{rgb}{0.745098,0.745098,0.745098}
  \definecolor{light grey}{rgb}{0.827451,0.827451,0.827451}
  \definecolor{LightGrey}{rgb}{0.827451,0.827451,0.827451}
  \definecolor{light gray}{rgb}{0.827451,0.827451,0.827451}
  \definecolor{LightGray}{rgb}{0.827451,0.827451,0.827451}
  \definecolor{midnight blue}{rgb}{0.098039,0.098039,0.439216}
  \definecolor{MidnightBlue}{rgb}{0.098039,0.098039,0.439216}
  \definecolor{navy}{rgb}{0.000000,0.000000,0.501961}
  \definecolor{navy blue}{rgb}{0.000000,0.000000,0.501961}
  \definecolor{NavyBlue}{rgb}{0.000000,0.000000,0.501961}
  \definecolor{cornflower blue}{rgb}{0.392157,0.584314,0.929412}
  \definecolor{CornflowerBlue}{rgb}{0.392157,0.584314,0.929412}
  \definecolor{dark slate blue}{rgb}{0.282353,0.239216,0.545098}
  \definecolor{DarkSlateBlue}{rgb}{0.282353,0.239216,0.545098}
  \definecolor{slate blue}{rgb}{0.415686,0.352941,0.803922}
  \definecolor{SlateBlue}{rgb}{0.415686,0.352941,0.803922}
  \definecolor{medium slate blue}{rgb}{0.482353,0.407843,0.933333}
  \definecolor{MediumSlateBlue}{rgb}{0.482353,0.407843,0.933333}
  \definecolor{light slate blue}{rgb}{0.517647,0.439216,1.000000}
  \definecolor{LightSlateBlue}{rgb}{0.517647,0.439216,1.000000}
  \definecolor{medium blue}{rgb}{0.000000,0.000000,0.803922}
  \definecolor{MediumBlue}{rgb}{0.000000,0.000000,0.803922}
  \definecolor{royal blue}{rgb}{0.254902,0.411765,0.882353}
  \definecolor{RoyalBlue}{rgb}{0.254902,0.411765,0.882353}
  \definecolor{blue}{rgb}{0.000000,0.000000,1.000000}
  \definecolor{dodger blue}{rgb}{0.117647,0.564706,1.000000}
  \definecolor{DodgerBlue}{rgb}{0.117647,0.564706,1.000000}
  \definecolor{deep sky blue}{rgb}{0.000000,0.749020,1.000000}
  \definecolor{DeepSkyBlue}{rgb}{0.000000,0.749020,1.000000}
  \definecolor{sky blue}{rgb}{0.529412,0.807843,0.921569}
  \definecolor{SkyBlue}{rgb}{0.529412,0.807843,0.921569}
  \definecolor{light sky blue}{rgb}{0.529412,0.807843,0.980392}
  \definecolor{LightSkyBlue}{rgb}{0.529412,0.807843,0.980392}
  \definecolor{steel blue}{rgb}{0.274510,0.509804,0.705882}
  \definecolor{SteelBlue}{rgb}{0.274510,0.509804,0.705882}
  \definecolor{light steel blue}{rgb}{0.690196,0.768627,0.870588}
  \definecolor{LightSteelBlue}{rgb}{0.690196,0.768627,0.870588}
  \definecolor{light blue}{rgb}{0.678431,0.847059,0.901961}
  \definecolor{LightBlue}{rgb}{0.678431,0.847059,0.901961}
  \definecolor{powder blue}{rgb}{0.690196,0.878431,0.901961}
  \definecolor{PowderBlue}{rgb}{0.690196,0.878431,0.901961}
  \definecolor{pale turquoise}{rgb}{0.686275,0.933333,0.933333}
  \definecolor{PaleTurquoise}{rgb}{0.686275,0.933333,0.933333}
  \definecolor{dark turquoise}{rgb}{0.000000,0.807843,0.819608}
  \definecolor{DarkTurquoise}{rgb}{0.000000,0.807843,0.819608}
  \definecolor{medium turquoise}{rgb}{0.282353,0.819608,0.800000}
  \definecolor{MediumTurquoise}{rgb}{0.282353,0.819608,0.800000}
  \definecolor{turquoise}{rgb}{0.250980,0.878431,0.815686}
  \definecolor{cyan}{rgb}{0.000000,1.000000,1.000000}
  \definecolor{light cyan}{rgb}{0.878431,1.000000,1.000000}
  \definecolor{LightCyan}{rgb}{0.878431,1.000000,1.000000}
  \definecolor{cadet blue}{rgb}{0.372549,0.619608,0.627451}
  \definecolor{CadetBlue}{rgb}{0.372549,0.619608,0.627451}
  \definecolor{medium aquamarine}{rgb}{0.400000,0.803922,0.666667}
  \definecolor{MediumAquamarine}{rgb}{0.400000,0.803922,0.666667}
  \definecolor{aquamarine}{rgb}{0.498039,1.000000,0.831373}
  \definecolor{dark green}{rgb}{0.000000,0.392157,0.000000}
  \definecolor{DarkGreen}{rgb}{0.000000,0.392157,0.000000}
  \definecolor{dark olive green}{rgb}{0.333333,0.419608,0.184314}
  \definecolor{DarkOliveGreen}{rgb}{0.333333,0.419608,0.184314}
  \definecolor{dark sea green}{rgb}{0.560784,0.737255,0.560784}
  \definecolor{DarkSeaGreen}{rgb}{0.560784,0.737255,0.560784}
  \definecolor{sea green}{rgb}{0.180392,0.545098,0.341176}
  \definecolor{SeaGreen}{rgb}{0.180392,0.545098,0.341176}
  \definecolor{medium sea green}{rgb}{0.235294,0.701961,0.443137}
  \definecolor{MediumSeaGreen}{rgb}{0.235294,0.701961,0.443137}
  \definecolor{light sea green}{rgb}{0.125490,0.698039,0.666667}
  \definecolor{LightSeaGreen}{rgb}{0.125490,0.698039,0.666667}
  \definecolor{pale green}{rgb}{0.596078,0.984314,0.596078}
  \definecolor{PaleGreen}{rgb}{0.596078,0.984314,0.596078}
  \definecolor{spring green}{rgb}{0.000000,1.000000,0.498039}
  \definecolor{SpringGreen}{rgb}{0.000000,1.000000,0.498039}
  \definecolor{lawn green}{rgb}{0.486275,0.988235,0.000000}
  \definecolor{LawnGreen}{rgb}{0.486275,0.988235,0.000000}
  \definecolor{green}{rgb}{0.000000,1.000000,0.000000}
  \definecolor{chartreuse}{rgb}{0.498039,1.000000,0.000000}
  \definecolor{medium spring green}{rgb}{0.000000,0.980392,0.603922}
  \definecolor{MediumSpringGreen}{rgb}{0.000000,0.980392,0.603922}
  \definecolor{green yellow}{rgb}{0.678431,1.000000,0.184314}
  \definecolor{GreenYellow}{rgb}{0.678431,1.000000,0.184314}
  \definecolor{lime green}{rgb}{0.196078,0.803922,0.196078}
  \definecolor{LimeGreen}{rgb}{0.196078,0.803922,0.196078}
  \definecolor{yellow green}{rgb}{0.603922,0.803922,0.196078}
  \definecolor{YellowGreen}{rgb}{0.603922,0.803922,0.196078}
  \definecolor{forest green}{rgb}{0.133333,0.545098,0.133333}
  \definecolor{ForestGreen}{rgb}{0.133333,0.545098,0.133333}
  \definecolor{olive drab}{rgb}{0.419608,0.556863,0.137255}
  \definecolor{OliveDrab}{rgb}{0.419608,0.556863,0.137255}
  \definecolor{dark khaki}{rgb}{0.741176,0.717647,0.419608}
  \definecolor{DarkKhaki}{rgb}{0.741176,0.717647,0.419608}
  \definecolor{khaki}{rgb}{0.941176,0.901961,0.549020}
  \definecolor{pale goldenrod}{rgb}{0.933333,0.909804,0.666667}
  \definecolor{PaleGoldenrod}{rgb}{0.933333,0.909804,0.666667}
  \definecolor{light goldenrod yellow}{rgb}{0.980392,0.980392,0.823529}
  \definecolor{LightGoldenrodYellow}{rgb}{0.980392,0.980392,0.823529}
  \definecolor{light yellow}{rgb}{1.000000,1.000000,0.878431}
  \definecolor{LightYellow}{rgb}{1.000000,1.000000,0.878431}
  \definecolor{yellow}{rgb}{1.000000,1.000000,0.000000}
  \definecolor{gold}{rgb}{1.000000,0.843137,0.000000}
  \definecolor{light goldenrod}{rgb}{0.933333,0.866667,0.509804}
  \definecolor{LightGoldenrod}{rgb}{0.933333,0.866667,0.509804}
  \definecolor{goldenrod}{rgb}{0.854902,0.647059,0.125490}
  \definecolor{dark goldenrod}{rgb}{0.721569,0.525490,0.043137}
  \definecolor{DarkGoldenrod}{rgb}{0.721569,0.525490,0.043137}
  \definecolor{rosy brown}{rgb}{0.737255,0.560784,0.560784}
  \definecolor{RosyBrown}{rgb}{0.737255,0.560784,0.560784}
  \definecolor{indian red}{rgb}{0.803922,0.360784,0.360784}
  \definecolor{IndianRed}{rgb}{0.803922,0.360784,0.360784}
  \definecolor{saddle brown}{rgb}{0.545098,0.270588,0.074510}
  \definecolor{SaddleBrown}{rgb}{0.545098,0.270588,0.074510}
  \definecolor{sienna}{rgb}{0.627451,0.321569,0.176471}
  \definecolor{peru}{rgb}{0.803922,0.521569,0.247059}
  \definecolor{burlywood}{rgb}{0.870588,0.721569,0.529412}
  \definecolor{beige}{rgb}{0.960784,0.960784,0.862745}
  \definecolor{wheat}{rgb}{0.960784,0.870588,0.701961}
  \definecolor{sandy brown}{rgb}{0.956863,0.643137,0.376471}
  \definecolor{SandyBrown}{rgb}{0.956863,0.643137,0.376471}
  \definecolor{tan}{rgb}{0.823529,0.705882,0.549020}
  \definecolor{chocolate}{rgb}{0.823529,0.411765,0.117647}
  \definecolor{firebrick}{rgb}{0.698039,0.133333,0.133333}
  \definecolor{brown}{rgb}{0.647059,0.164706,0.164706}
  \definecolor{dark salmon}{rgb}{0.913725,0.588235,0.478431}
  \definecolor{DarkSalmon}{rgb}{0.913725,0.588235,0.478431}
  \definecolor{salmon}{rgb}{0.980392,0.501961,0.447059}
  \definecolor{light salmon}{rgb}{1.000000,0.627451,0.478431}
  \definecolor{LightSalmon}{rgb}{1.000000,0.627451,0.478431}
  \definecolor{orange}{rgb}{1.000000,0.647059,0.000000}
  \definecolor{dark orange}{rgb}{1.000000,0.549020,0.000000}
  \definecolor{DarkOrange}{rgb}{1.000000,0.549020,0.000000}
  \definecolor{coral}{rgb}{1.000000,0.498039,0.313726}
  \definecolor{light coral}{rgb}{0.941176,0.501961,0.501961}
  \definecolor{LightCoral}{rgb}{0.941176,0.501961,0.501961}
  \definecolor{tomato}{rgb}{1.000000,0.388235,0.278431}
  \definecolor{orange red}{rgb}{1.000000,0.270588,0.000000}
  \definecolor{OrangeRed}{rgb}{1.000000,0.270588,0.000000}
  \definecolor{red}{rgb}{1.000000,0.000000,0.000000}
  \definecolor{hot pink}{rgb}{1.000000,0.411765,0.705882}
  \definecolor{HotPink}{rgb}{1.000000,0.411765,0.705882}
  \definecolor{deep pink}{rgb}{1.000000,0.078431,0.576471}
  \definecolor{DeepPink}{rgb}{1.000000,0.078431,0.576471}
  \definecolor{pink}{rgb}{1.000000,0.752941,0.796078}
  \definecolor{light pink}{rgb}{1.000000,0.713726,0.756863}
  \definecolor{LightPink}{rgb}{1.000000,0.713726,0.756863}
  \definecolor{pale violet red}{rgb}{0.858824,0.439216,0.576471}
  \definecolor{PaleVioletRed}{rgb}{0.858824,0.439216,0.576471}
  \definecolor{maroon}{rgb}{0.690196,0.188235,0.376471}
  \definecolor{medium violet red}{rgb}{0.780392,0.082353,0.521569}
  \definecolor{MediumVioletRed}{rgb}{0.780392,0.082353,0.521569}
  \definecolor{violet red}{rgb}{0.815686,0.125490,0.564706}
  \definecolor{VioletRed}{rgb}{0.815686,0.125490,0.564706}
  \definecolor{magenta}{rgb}{1.000000,0.000000,1.000000}
  \definecolor{violet}{rgb}{0.933333,0.509804,0.933333}
  \definecolor{plum}{rgb}{0.866667,0.627451,0.866667}
  \definecolor{orchid}{rgb}{0.854902,0.439216,0.839216}
  \definecolor{medium orchid}{rgb}{0.729412,0.333333,0.827451}
  \definecolor{MediumOrchid}{rgb}{0.729412,0.333333,0.827451}
  \definecolor{dark orchid}{rgb}{0.600000,0.196078,0.800000}
  \definecolor{DarkOrchid}{rgb}{0.600000,0.196078,0.800000}
  \definecolor{dark violet}{rgb}{0.580392,0.000000,0.827451}
  \definecolor{DarkViolet}{rgb}{0.580392,0.000000,0.827451}
  \definecolor{blue violet}{rgb}{0.541176,0.168627,0.886275}
  \definecolor{BlueViolet}{rgb}{0.541176,0.168627,0.886275}
  \definecolor{purple}{rgb}{0.627451,0.125490,0.941176}
  \definecolor{medium purple}{rgb}{0.576471,0.439216,0.858824}
  \definecolor{MediumPurple}{rgb}{0.576471,0.439216,0.858824}
  \definecolor{thistle}{rgb}{0.847059,0.749020,0.847059}
  \definecolor{snow1}{rgb}{1.000000,0.980392,0.980392}
  \definecolor{snow2}{rgb}{0.933333,0.913725,0.913725}
  \definecolor{snow3}{rgb}{0.803922,0.788235,0.788235}
  \definecolor{snow4}{rgb}{0.545098,0.537255,0.537255}
  \definecolor{seashell1}{rgb}{1.000000,0.960784,0.933333}
  \definecolor{seashell2}{rgb}{0.933333,0.898039,0.870588}
  \definecolor{seashell3}{rgb}{0.803922,0.772549,0.749020}
  \definecolor{seashell4}{rgb}{0.545098,0.525490,0.509804}
  \definecolor{AntiqueWhite1}{rgb}{1.000000,0.937255,0.858824}
  \definecolor{AntiqueWhite2}{rgb}{0.933333,0.874510,0.800000}
  \definecolor{AntiqueWhite3}{rgb}{0.803922,0.752941,0.690196}
  \definecolor{AntiqueWhite4}{rgb}{0.545098,0.513726,0.470588}
  \definecolor{bisque1}{rgb}{1.000000,0.894118,0.768627}
  \definecolor{bisque2}{rgb}{0.933333,0.835294,0.717647}
  \definecolor{bisque3}{rgb}{0.803922,0.717647,0.619608}
  \definecolor{bisque4}{rgb}{0.545098,0.490196,0.419608}
  \definecolor{PeachPuff1}{rgb}{1.000000,0.854902,0.725490}
  \definecolor{PeachPuff2}{rgb}{0.933333,0.796078,0.678431}
  \definecolor{PeachPuff3}{rgb}{0.803922,0.686275,0.584314}
  \definecolor{PeachPuff4}{rgb}{0.545098,0.466667,0.396078}
  \definecolor{NavajoWhite1}{rgb}{1.000000,0.870588,0.678431}
  \definecolor{NavajoWhite2}{rgb}{0.933333,0.811765,0.631373}
  \definecolor{NavajoWhite3}{rgb}{0.803922,0.701961,0.545098}
  \definecolor{NavajoWhite4}{rgb}{0.545098,0.474510,0.368627}
  \definecolor{LemonChiffon1}{rgb}{1.000000,0.980392,0.803922}
  \definecolor{LemonChiffon2}{rgb}{0.933333,0.913725,0.749020}
  \definecolor{LemonChiffon3}{rgb}{0.803922,0.788235,0.647059}
  \definecolor{LemonChiffon4}{rgb}{0.545098,0.537255,0.439216}
  \definecolor{cornsilk1}{rgb}{1.000000,0.972549,0.862745}
  \definecolor{cornsilk2}{rgb}{0.933333,0.909804,0.803922}
  \definecolor{cornsilk3}{rgb}{0.803922,0.784314,0.694118}
  \definecolor{cornsilk4}{rgb}{0.545098,0.533333,0.470588}
  \definecolor{ivory1}{rgb}{1.000000,1.000000,0.941176}
  \definecolor{ivory2}{rgb}{0.933333,0.933333,0.878431}
  \definecolor{ivory3}{rgb}{0.803922,0.803922,0.756863}
  \definecolor{ivory4}{rgb}{0.545098,0.545098,0.513726}
  \definecolor{honeydew1}{rgb}{0.941176,1.000000,0.941176}
  \definecolor{honeydew2}{rgb}{0.878431,0.933333,0.878431}
  \definecolor{honeydew3}{rgb}{0.756863,0.803922,0.756863}
  \definecolor{honeydew4}{rgb}{0.513726,0.545098,0.513726}
  \definecolor{LavenderBlush1}{rgb}{1.000000,0.941176,0.960784}
  \definecolor{LavenderBlush2}{rgb}{0.933333,0.878431,0.898039}
  \definecolor{LavenderBlush3}{rgb}{0.803922,0.756863,0.772549}
  \definecolor{LavenderBlush4}{rgb}{0.545098,0.513726,0.525490}
  \definecolor{MistyRose1}{rgb}{1.000000,0.894118,0.882353}
  \definecolor{MistyRose2}{rgb}{0.933333,0.835294,0.823529}
  \definecolor{MistyRose3}{rgb}{0.803922,0.717647,0.709804}
  \definecolor{MistyRose4}{rgb}{0.545098,0.490196,0.482353}
  \definecolor{azure1}{rgb}{0.941176,1.000000,1.000000}
  \definecolor{azure2}{rgb}{0.878431,0.933333,0.933333}
  \definecolor{azure3}{rgb}{0.756863,0.803922,0.803922}
  \definecolor{azure4}{rgb}{0.513726,0.545098,0.545098}
  \definecolor{SlateBlue1}{rgb}{0.513726,0.435294,1.000000}
  \definecolor{SlateBlue2}{rgb}{0.478431,0.403922,0.933333}
  \definecolor{SlateBlue3}{rgb}{0.411765,0.349020,0.803922}
  \definecolor{SlateBlue4}{rgb}{0.278431,0.235294,0.545098}
  \definecolor{RoyalBlue1}{rgb}{0.282353,0.462745,1.000000}
  \definecolor{RoyalBlue2}{rgb}{0.262745,0.431373,0.933333}
  \definecolor{RoyalBlue3}{rgb}{0.227451,0.372549,0.803922}
  \definecolor{RoyalBlue4}{rgb}{0.152941,0.250980,0.545098}
  \definecolor{blue1}{rgb}{0.000000,0.000000,1.000000}
  \definecolor{blue2}{rgb}{0.000000,0.000000,0.933333}
  \definecolor{blue3}{rgb}{0.000000,0.000000,0.803922}
  \definecolor{blue4}{rgb}{0.000000,0.000000,0.545098}
  \definecolor{DodgerBlue1}{rgb}{0.117647,0.564706,1.000000}
  \definecolor{DodgerBlue2}{rgb}{0.109804,0.525490,0.933333}
  \definecolor{DodgerBlue3}{rgb}{0.094118,0.454902,0.803922}
  \definecolor{DodgerBlue4}{rgb}{0.062745,0.305882,0.545098}
  \definecolor{SteelBlue1}{rgb}{0.388235,0.721569,1.000000}
  \definecolor{SteelBlue2}{rgb}{0.360784,0.674510,0.933333}
  \definecolor{SteelBlue3}{rgb}{0.309804,0.580392,0.803922}
  \definecolor{SteelBlue4}{rgb}{0.211765,0.392157,0.545098}
  \definecolor{DeepSkyBlue1}{rgb}{0.000000,0.749020,1.000000}
  \definecolor{DeepSkyBlue2}{rgb}{0.000000,0.698039,0.933333}
  \definecolor{DeepSkyBlue3}{rgb}{0.000000,0.603922,0.803922}
  \definecolor{DeepSkyBlue4}{rgb}{0.000000,0.407843,0.545098}
  \definecolor{SkyBlue1}{rgb}{0.529412,0.807843,1.000000}
  \definecolor{SkyBlue2}{rgb}{0.494118,0.752941,0.933333}
  \definecolor{SkyBlue3}{rgb}{0.423529,0.650980,0.803922}
  \definecolor{SkyBlue4}{rgb}{0.290196,0.439216,0.545098}
  \definecolor{LightSkyBlue1}{rgb}{0.690196,0.886275,1.000000}
  \definecolor{LightSkyBlue2}{rgb}{0.643137,0.827451,0.933333}
  \definecolor{LightSkyBlue3}{rgb}{0.552941,0.713726,0.803922}
  \definecolor{LightSkyBlue4}{rgb}{0.376471,0.482353,0.545098}
  \definecolor{SlateGray1}{rgb}{0.776471,0.886275,1.000000}
  \definecolor{SlateGray2}{rgb}{0.725490,0.827451,0.933333}
  \definecolor{SlateGray3}{rgb}{0.623529,0.713726,0.803922}
  \definecolor{SlateGray4}{rgb}{0.423529,0.482353,0.545098}
  \definecolor{LightSteelBlue1}{rgb}{0.792157,0.882353,1.000000}
  \definecolor{LightSteelBlue2}{rgb}{0.737255,0.823529,0.933333}
  \definecolor{LightSteelBlue3}{rgb}{0.635294,0.709804,0.803922}
  \definecolor{LightSteelBlue4}{rgb}{0.431373,0.482353,0.545098}
  \definecolor{LightBlue1}{rgb}{0.749020,0.937255,1.000000}
  \definecolor{LightBlue2}{rgb}{0.698039,0.874510,0.933333}
  \definecolor{LightBlue3}{rgb}{0.603922,0.752941,0.803922}
  \definecolor{LightBlue4}{rgb}{0.407843,0.513726,0.545098}
  \definecolor{LightCyan1}{rgb}{0.878431,1.000000,1.000000}
  \definecolor{LightCyan2}{rgb}{0.819608,0.933333,0.933333}
  \definecolor{LightCyan3}{rgb}{0.705882,0.803922,0.803922}
  \definecolor{LightCyan4}{rgb}{0.478431,0.545098,0.545098}
  \definecolor{PaleTurquoise1}{rgb}{0.733333,1.000000,1.000000}
  \definecolor{PaleTurquoise2}{rgb}{0.682353,0.933333,0.933333}
  \definecolor{PaleTurquoise3}{rgb}{0.588235,0.803922,0.803922}
  \definecolor{PaleTurquoise4}{rgb}{0.400000,0.545098,0.545098}
  \definecolor{CadetBlue1}{rgb}{0.596078,0.960784,1.000000}
  \definecolor{CadetBlue2}{rgb}{0.556863,0.898039,0.933333}
  \definecolor{CadetBlue3}{rgb}{0.478431,0.772549,0.803922}
  \definecolor{CadetBlue4}{rgb}{0.325490,0.525490,0.545098}
  \definecolor{turquoise1}{rgb}{0.000000,0.960784,1.000000}
  \definecolor{turquoise2}{rgb}{0.000000,0.898039,0.933333}
  \definecolor{turquoise3}{rgb}{0.000000,0.772549,0.803922}
  \definecolor{turquoise4}{rgb}{0.000000,0.525490,0.545098}
  \definecolor{cyan1}{rgb}{0.000000,1.000000,1.000000}
  \definecolor{cyan2}{rgb}{0.000000,0.933333,0.933333}
  \definecolor{cyan3}{rgb}{0.000000,0.803922,0.803922}
  \definecolor{cyan4}{rgb}{0.000000,0.545098,0.545098}
  \definecolor{DarkSlateGray1}{rgb}{0.592157,1.000000,1.000000}
  \definecolor{DarkSlateGray2}{rgb}{0.552941,0.933333,0.933333}
  \definecolor{DarkSlateGray3}{rgb}{0.474510,0.803922,0.803922}
  \definecolor{DarkSlateGray4}{rgb}{0.321569,0.545098,0.545098}
  \definecolor{aquamarine1}{rgb}{0.498039,1.000000,0.831373}
  \definecolor{aquamarine2}{rgb}{0.462745,0.933333,0.776471}
  \definecolor{aquamarine3}{rgb}{0.400000,0.803922,0.666667}
  \definecolor{aquamarine4}{rgb}{0.270588,0.545098,0.454902}
  \definecolor{DarkSeaGreen1}{rgb}{0.756863,1.000000,0.756863}
  \definecolor{DarkSeaGreen2}{rgb}{0.705882,0.933333,0.705882}
  \definecolor{DarkSeaGreen3}{rgb}{0.607843,0.803922,0.607843}
  \definecolor{DarkSeaGreen4}{rgb}{0.411765,0.545098,0.411765}
  \definecolor{SeaGreen1}{rgb}{0.329412,1.000000,0.623529}
  \definecolor{SeaGreen2}{rgb}{0.305882,0.933333,0.580392}
  \definecolor{SeaGreen3}{rgb}{0.262745,0.803922,0.501961}
  \definecolor{SeaGreen4}{rgb}{0.180392,0.545098,0.341176}
  \definecolor{PaleGreen1}{rgb}{0.603922,1.000000,0.603922}
  \definecolor{PaleGreen2}{rgb}{0.564706,0.933333,0.564706}
  \definecolor{PaleGreen3}{rgb}{0.486275,0.803922,0.486275}
  \definecolor{PaleGreen4}{rgb}{0.329412,0.545098,0.329412}
  \definecolor{SpringGreen1}{rgb}{0.000000,1.000000,0.498039}
  \definecolor{SpringGreen2}{rgb}{0.000000,0.933333,0.462745}
  \definecolor{SpringGreen3}{rgb}{0.000000,0.803922,0.400000}
  \definecolor{SpringGreen4}{rgb}{0.000000,0.545098,0.270588}
  \definecolor{green1}{rgb}{0.000000,1.000000,0.000000}
  \definecolor{green2}{rgb}{0.000000,0.933333,0.000000}
  \definecolor{green3}{rgb}{0.000000,0.803922,0.000000}
  \definecolor{green4}{rgb}{0.000000,0.545098,0.000000}
  \definecolor{chartreuse1}{rgb}{0.498039,1.000000,0.000000}
  \definecolor{chartreuse2}{rgb}{0.462745,0.933333,0.000000}
  \definecolor{chartreuse3}{rgb}{0.400000,0.803922,0.000000}
  \definecolor{chartreuse4}{rgb}{0.270588,0.545098,0.000000}
  \definecolor{OliveDrab1}{rgb}{0.752941,1.000000,0.243137}
  \definecolor{OliveDrab2}{rgb}{0.701961,0.933333,0.227451}
  \definecolor{OliveDrab3}{rgb}{0.603922,0.803922,0.196078}
  \definecolor{OliveDrab4}{rgb}{0.411765,0.545098,0.133333}
  \definecolor{DarkOliveGreen1}{rgb}{0.792157,1.000000,0.439216}
  \definecolor{DarkOliveGreen2}{rgb}{0.737255,0.933333,0.407843}
  \definecolor{DarkOliveGreen3}{rgb}{0.635294,0.803922,0.352941}
  \definecolor{DarkOliveGreen4}{rgb}{0.431373,0.545098,0.239216}
  \definecolor{khaki1}{rgb}{1.000000,0.964706,0.560784}
  \definecolor{khaki2}{rgb}{0.933333,0.901961,0.521569}
  \definecolor{khaki3}{rgb}{0.803922,0.776471,0.450980}
  \definecolor{khaki4}{rgb}{0.545098,0.525490,0.305882}
  \definecolor{LightGoldenrod1}{rgb}{1.000000,0.925490,0.545098}
  \definecolor{LightGoldenrod2}{rgb}{0.933333,0.862745,0.509804}
  \definecolor{LightGoldenrod3}{rgb}{0.803922,0.745098,0.439216}
  \definecolor{LightGoldenrod4}{rgb}{0.545098,0.505882,0.298039}
  \definecolor{LightYellow1}{rgb}{1.000000,1.000000,0.878431}
  \definecolor{LightYellow2}{rgb}{0.933333,0.933333,0.819608}
  \definecolor{LightYellow3}{rgb}{0.803922,0.803922,0.705882}
  \definecolor{LightYellow4}{rgb}{0.545098,0.545098,0.478431}
  \definecolor{yellow1}{rgb}{1.000000,1.000000,0.000000}
  \definecolor{yellow2}{rgb}{0.933333,0.933333,0.000000}
  \definecolor{yellow3}{rgb}{0.803922,0.803922,0.000000}
  \definecolor{yellow4}{rgb}{0.545098,0.545098,0.000000}
  \definecolor{gold1}{rgb}{1.000000,0.843137,0.000000}
  \definecolor{gold2}{rgb}{0.933333,0.788235,0.000000}
  \definecolor{gold3}{rgb}{0.803922,0.678431,0.000000}
  \definecolor{gold4}{rgb}{0.545098,0.458824,0.000000}
  \definecolor{goldenrod1}{rgb}{1.000000,0.756863,0.145098}
  \definecolor{goldenrod2}{rgb}{0.933333,0.705882,0.133333}
  \definecolor{goldenrod3}{rgb}{0.803922,0.607843,0.113725}
  \definecolor{goldenrod4}{rgb}{0.545098,0.411765,0.078431}
  \definecolor{DarkGoldenrod1}{rgb}{1.000000,0.725490,0.058824}
  \definecolor{DarkGoldenrod2}{rgb}{0.933333,0.678431,0.054902}
  \definecolor{DarkGoldenrod3}{rgb}{0.803922,0.584314,0.047059}
  \definecolor{DarkGoldenrod4}{rgb}{0.545098,0.396078,0.031373}
  \definecolor{RosyBrown1}{rgb}{1.000000,0.756863,0.756863}
  \definecolor{RosyBrown2}{rgb}{0.933333,0.705882,0.705882}
  \definecolor{RosyBrown3}{rgb}{0.803922,0.607843,0.607843}
  \definecolor{RosyBrown4}{rgb}{0.545098,0.411765,0.411765}
  \definecolor{IndianRed1}{rgb}{1.000000,0.415686,0.415686}
  \definecolor{IndianRed2}{rgb}{0.933333,0.388235,0.388235}
  \definecolor{IndianRed3}{rgb}{0.803922,0.333333,0.333333}
  \definecolor{IndianRed4}{rgb}{0.545098,0.227451,0.227451}
  \definecolor{sienna1}{rgb}{1.000000,0.509804,0.278431}
  \definecolor{sienna2}{rgb}{0.933333,0.474510,0.258824}
  \definecolor{sienna3}{rgb}{0.803922,0.407843,0.223529}
  \definecolor{sienna4}{rgb}{0.545098,0.278431,0.149020}
  \definecolor{burlywood1}{rgb}{1.000000,0.827451,0.607843}
  \definecolor{burlywood2}{rgb}{0.933333,0.772549,0.568627}
  \definecolor{burlywood3}{rgb}{0.803922,0.666667,0.490196}
  \definecolor{burlywood4}{rgb}{0.545098,0.450980,0.333333}
  \definecolor{wheat1}{rgb}{1.000000,0.905882,0.729412}
  \definecolor{wheat2}{rgb}{0.933333,0.847059,0.682353}
  \definecolor{wheat3}{rgb}{0.803922,0.729412,0.588235}
  \definecolor{wheat4}{rgb}{0.545098,0.494118,0.400000}
  \definecolor{tan1}{rgb}{1.000000,0.647059,0.309804}
  \definecolor{tan2}{rgb}{0.933333,0.603922,0.286275}
  \definecolor{tan3}{rgb}{0.803922,0.521569,0.247059}
  \definecolor{tan4}{rgb}{0.545098,0.352941,0.168627}
  \definecolor{chocolate1}{rgb}{1.000000,0.498039,0.141176}
  \definecolor{chocolate2}{rgb}{0.933333,0.462745,0.129412}
  \definecolor{chocolate3}{rgb}{0.803922,0.400000,0.113725}
  \definecolor{chocolate4}{rgb}{0.545098,0.270588,0.074510}
  \definecolor{firebrick1}{rgb}{1.000000,0.188235,0.188235}
  \definecolor{firebrick2}{rgb}{0.933333,0.172549,0.172549}
  \definecolor{firebrick3}{rgb}{0.803922,0.149020,0.149020}
  \definecolor{firebrick4}{rgb}{0.545098,0.101961,0.101961}
  \definecolor{brown1}{rgb}{1.000000,0.250980,0.250980}
  \definecolor{brown2}{rgb}{0.933333,0.231373,0.231373}
  \definecolor{brown3}{rgb}{0.803922,0.200000,0.200000}
  \definecolor{brown4}{rgb}{0.545098,0.137255,0.137255}
  \definecolor{salmon1}{rgb}{1.000000,0.549020,0.411765}
  \definecolor{salmon2}{rgb}{0.933333,0.509804,0.384314}
  \definecolor{salmon3}{rgb}{0.803922,0.439216,0.329412}
  \definecolor{salmon4}{rgb}{0.545098,0.298039,0.223529}
  \definecolor{LightSalmon1}{rgb}{1.000000,0.627451,0.478431}
  \definecolor{LightSalmon2}{rgb}{0.933333,0.584314,0.447059}
  \definecolor{LightSalmon3}{rgb}{0.803922,0.505882,0.384314}
  \definecolor{LightSalmon4}{rgb}{0.545098,0.341176,0.258824}
  \definecolor{orange1}{rgb}{1.000000,0.647059,0.000000}
  \definecolor{orange2}{rgb}{0.933333,0.603922,0.000000}
  \definecolor{orange3}{rgb}{0.803922,0.521569,0.000000}
  \definecolor{orange4}{rgb}{0.545098,0.352941,0.000000}
  \definecolor{DarkOrange1}{rgb}{1.000000,0.498039,0.000000}
  \definecolor{DarkOrange2}{rgb}{0.933333,0.462745,0.000000}
  \definecolor{DarkOrange3}{rgb}{0.803922,0.400000,0.000000}
  \definecolor{DarkOrange4}{rgb}{0.545098,0.270588,0.000000}
  \definecolor{coral1}{rgb}{1.000000,0.447059,0.337255}
  \definecolor{coral2}{rgb}{0.933333,0.415686,0.313726}
  \definecolor{coral3}{rgb}{0.803922,0.356863,0.270588}
  \definecolor{coral4}{rgb}{0.545098,0.243137,0.184314}
  \definecolor{tomato1}{rgb}{1.000000,0.388235,0.278431}
  \definecolor{tomato2}{rgb}{0.933333,0.360784,0.258824}
  \definecolor{tomato3}{rgb}{0.803922,0.309804,0.223529}
  \definecolor{tomato4}{rgb}{0.545098,0.211765,0.149020}
  \definecolor{OrangeRed1}{rgb}{1.000000,0.270588,0.000000}
  \definecolor{OrangeRed2}{rgb}{0.933333,0.250980,0.000000}
  \definecolor{OrangeRed3}{rgb}{0.803922,0.215686,0.000000}
  \definecolor{OrangeRed4}{rgb}{0.545098,0.145098,0.000000}
  \definecolor{red1}{rgb}{1.000000,0.000000,0.000000}
  \definecolor{red2}{rgb}{0.933333,0.000000,0.000000}
  \definecolor{red3}{rgb}{0.803922,0.000000,0.000000}
  \definecolor{red4}{rgb}{0.545098,0.000000,0.000000}
  \definecolor{DeepPink1}{rgb}{1.000000,0.078431,0.576471}
  \definecolor{DeepPink2}{rgb}{0.933333,0.070588,0.537255}
  \definecolor{DeepPink3}{rgb}{0.803922,0.062745,0.462745}
  \definecolor{DeepPink4}{rgb}{0.545098,0.039216,0.313726}
  \definecolor{HotPink1}{rgb}{1.000000,0.431373,0.705882}
  \definecolor{HotPink2}{rgb}{0.933333,0.415686,0.654902}
  \definecolor{HotPink3}{rgb}{0.803922,0.376471,0.564706}
  \definecolor{HotPink4}{rgb}{0.545098,0.227451,0.384314}
  \definecolor{pink1}{rgb}{1.000000,0.709804,0.772549}
  \definecolor{pink2}{rgb}{0.933333,0.662745,0.721569}
  \definecolor{pink3}{rgb}{0.803922,0.568627,0.619608}
  \definecolor{pink4}{rgb}{0.545098,0.388235,0.423529}
  \definecolor{LightPink1}{rgb}{1.000000,0.682353,0.725490}
  \definecolor{LightPink2}{rgb}{0.933333,0.635294,0.678431}
  \definecolor{LightPink3}{rgb}{0.803922,0.549020,0.584314}
  \definecolor{LightPink4}{rgb}{0.545098,0.372549,0.396078}
  \definecolor{PaleVioletRed1}{rgb}{1.000000,0.509804,0.670588}
  \definecolor{PaleVioletRed2}{rgb}{0.933333,0.474510,0.623529}
  \definecolor{PaleVioletRed3}{rgb}{0.803922,0.407843,0.537255}
  \definecolor{PaleVioletRed4}{rgb}{0.545098,0.278431,0.364706}
  \definecolor{maroon1}{rgb}{1.000000,0.203922,0.701961}
  \definecolor{maroon2}{rgb}{0.933333,0.188235,0.654902}
  \definecolor{maroon3}{rgb}{0.803922,0.160784,0.564706}
  \definecolor{maroon4}{rgb}{0.545098,0.109804,0.384314}
  \definecolor{VioletRed1}{rgb}{1.000000,0.243137,0.588235}
  \definecolor{VioletRed2}{rgb}{0.933333,0.227451,0.549020}
  \definecolor{VioletRed3}{rgb}{0.803922,0.196078,0.470588}
  \definecolor{VioletRed4}{rgb}{0.545098,0.133333,0.321569}
  \definecolor{magenta1}{rgb}{1.000000,0.000000,1.000000}
  \definecolor{magenta2}{rgb}{0.933333,0.000000,0.933333}
  \definecolor{magenta3}{rgb}{0.803922,0.000000,0.803922}
  \definecolor{magenta4}{rgb}{0.545098,0.000000,0.545098}
  \definecolor{orchid1}{rgb}{1.000000,0.513726,0.980392}
  \definecolor{orchid2}{rgb}{0.933333,0.478431,0.913725}
  \definecolor{orchid3}{rgb}{0.803922,0.411765,0.788235}
  \definecolor{orchid4}{rgb}{0.545098,0.278431,0.537255}
  \definecolor{plum1}{rgb}{1.000000,0.733333,1.000000}
  \definecolor{plum2}{rgb}{0.933333,0.682353,0.933333}
  \definecolor{plum3}{rgb}{0.803922,0.588235,0.803922}
  \definecolor{plum4}{rgb}{0.545098,0.400000,0.545098}
  \definecolor{MediumOrchid1}{rgb}{0.878431,0.400000,1.000000}
  \definecolor{MediumOrchid2}{rgb}{0.819608,0.372549,0.933333}
  \definecolor{MediumOrchid3}{rgb}{0.705882,0.321569,0.803922}
  \definecolor{MediumOrchid4}{rgb}{0.478431,0.215686,0.545098}
  \definecolor{DarkOrchid1}{rgb}{0.749020,0.243137,1.000000}
  \definecolor{DarkOrchid2}{rgb}{0.698039,0.227451,0.933333}
  \definecolor{DarkOrchid3}{rgb}{0.603922,0.196078,0.803922}
  \definecolor{DarkOrchid4}{rgb}{0.407843,0.133333,0.545098}
  \definecolor{purple1}{rgb}{0.607843,0.188235,1.000000}
  \definecolor{purple2}{rgb}{0.568627,0.172549,0.933333}
  \definecolor{purple3}{rgb}{0.490196,0.149020,0.803922}
  \definecolor{purple4}{rgb}{0.333333,0.101961,0.545098}
  \definecolor{MediumPurple1}{rgb}{0.670588,0.509804,1.000000}
  \definecolor{MediumPurple2}{rgb}{0.623529,0.474510,0.933333}
  \definecolor{MediumPurple3}{rgb}{0.537255,0.407843,0.803922}
  \definecolor{MediumPurple4}{rgb}{0.364706,0.278431,0.545098}
  \definecolor{thistle1}{rgb}{1.000000,0.882353,1.000000}
  \definecolor{thistle2}{rgb}{0.933333,0.823529,0.933333}
  \definecolor{thistle3}{rgb}{0.803922,0.709804,0.803922}
  \definecolor{thistle4}{rgb}{0.545098,0.482353,0.545098}
  \definecolor{gray0}{rgb}{0.000000,0.000000,0.000000}
  \definecolor{grey0}{rgb}{0.000000,0.000000,0.000000}
  \definecolor{gray1}{rgb}{0.011765,0.011765,0.011765}
  \definecolor{grey1}{rgb}{0.011765,0.011765,0.011765}
  \definecolor{gray2}{rgb}{0.019608,0.019608,0.019608}
  \definecolor{grey2}{rgb}{0.019608,0.019608,0.019608}
  \definecolor{gray3}{rgb}{0.031373,0.031373,0.031373}
  \definecolor{grey3}{rgb}{0.031373,0.031373,0.031373}
  \definecolor{gray4}{rgb}{0.039216,0.039216,0.039216}
  \definecolor{grey4}{rgb}{0.039216,0.039216,0.039216}
  \definecolor{gray5}{rgb}{0.050980,0.050980,0.050980}
  \definecolor{grey5}{rgb}{0.050980,0.050980,0.050980}
  \definecolor{gray6}{rgb}{0.058824,0.058824,0.058824}
  \definecolor{grey6}{rgb}{0.058824,0.058824,0.058824}
  \definecolor{gray7}{rgb}{0.070588,0.070588,0.070588}
  \definecolor{grey7}{rgb}{0.070588,0.070588,0.070588}
  \definecolor{gray8}{rgb}{0.078431,0.078431,0.078431}
  \definecolor{grey8}{rgb}{0.078431,0.078431,0.078431}
  \definecolor{gray9}{rgb}{0.090196,0.090196,0.090196}
  \definecolor{grey9}{rgb}{0.090196,0.090196,0.090196}
  \definecolor{gray10}{rgb}{0.101961,0.101961,0.101961}
  \definecolor{grey10}{rgb}{0.101961,0.101961,0.101961}
  \definecolor{gray11}{rgb}{0.109804,0.109804,0.109804}
  \definecolor{grey11}{rgb}{0.109804,0.109804,0.109804}
  \definecolor{gray12}{rgb}{0.121569,0.121569,0.121569}
  \definecolor{grey12}{rgb}{0.121569,0.121569,0.121569}
  \definecolor{gray13}{rgb}{0.129412,0.129412,0.129412}
  \definecolor{grey13}{rgb}{0.129412,0.129412,0.129412}
  \definecolor{gray14}{rgb}{0.141176,0.141176,0.141176}
  \definecolor{grey14}{rgb}{0.141176,0.141176,0.141176}
  \definecolor{gray15}{rgb}{0.149020,0.149020,0.149020}
  \definecolor{grey15}{rgb}{0.149020,0.149020,0.149020}
  \definecolor{gray16}{rgb}{0.160784,0.160784,0.160784}
  \definecolor{grey16}{rgb}{0.160784,0.160784,0.160784}
  \definecolor{gray17}{rgb}{0.168627,0.168627,0.168627}
  \definecolor{grey17}{rgb}{0.168627,0.168627,0.168627}
  \definecolor{gray18}{rgb}{0.180392,0.180392,0.180392}
  \definecolor{grey18}{rgb}{0.180392,0.180392,0.180392}
  \definecolor{gray19}{rgb}{0.188235,0.188235,0.188235}
  \definecolor{grey19}{rgb}{0.188235,0.188235,0.188235}
  \definecolor{gray20}{rgb}{0.200000,0.200000,0.200000}
  \definecolor{grey20}{rgb}{0.200000,0.200000,0.200000}
  \definecolor{gray21}{rgb}{0.211765,0.211765,0.211765}
  \definecolor{grey21}{rgb}{0.211765,0.211765,0.211765}
  \definecolor{gray22}{rgb}{0.219608,0.219608,0.219608}
  \definecolor{grey22}{rgb}{0.219608,0.219608,0.219608}
  \definecolor{gray23}{rgb}{0.231373,0.231373,0.231373}
  \definecolor{grey23}{rgb}{0.231373,0.231373,0.231373}
  \definecolor{gray24}{rgb}{0.239216,0.239216,0.239216}
  \definecolor{grey24}{rgb}{0.239216,0.239216,0.239216}
  \definecolor{gray25}{rgb}{0.250980,0.250980,0.250980}
  \definecolor{grey25}{rgb}{0.250980,0.250980,0.250980}
  \definecolor{gray26}{rgb}{0.258824,0.258824,0.258824}
  \definecolor{grey26}{rgb}{0.258824,0.258824,0.258824}
  \definecolor{gray27}{rgb}{0.270588,0.270588,0.270588}
  \definecolor{grey27}{rgb}{0.270588,0.270588,0.270588}
  \definecolor{gray28}{rgb}{0.278431,0.278431,0.278431}
  \definecolor{grey28}{rgb}{0.278431,0.278431,0.278431}
  \definecolor{gray29}{rgb}{0.290196,0.290196,0.290196}
  \definecolor{grey29}{rgb}{0.290196,0.290196,0.290196}
  \definecolor{gray30}{rgb}{0.301961,0.301961,0.301961}
  \definecolor{grey30}{rgb}{0.301961,0.301961,0.301961}
  \definecolor{gray31}{rgb}{0.309804,0.309804,0.309804}
  \definecolor{grey31}{rgb}{0.309804,0.309804,0.309804}
  \definecolor{gray32}{rgb}{0.321569,0.321569,0.321569}
  \definecolor{grey32}{rgb}{0.321569,0.321569,0.321569}
  \definecolor{gray33}{rgb}{0.329412,0.329412,0.329412}
  \definecolor{grey33}{rgb}{0.329412,0.329412,0.329412}
  \definecolor{gray34}{rgb}{0.341176,0.341176,0.341176}
  \definecolor{grey34}{rgb}{0.341176,0.341176,0.341176}
  \definecolor{gray35}{rgb}{0.349020,0.349020,0.349020}
  \definecolor{grey35}{rgb}{0.349020,0.349020,0.349020}
  \definecolor{gray36}{rgb}{0.360784,0.360784,0.360784}
  \definecolor{grey36}{rgb}{0.360784,0.360784,0.360784}
  \definecolor{gray37}{rgb}{0.368627,0.368627,0.368627}
  \definecolor{grey37}{rgb}{0.368627,0.368627,0.368627}
  \definecolor{gray38}{rgb}{0.380392,0.380392,0.380392}
  \definecolor{grey38}{rgb}{0.380392,0.380392,0.380392}
  \definecolor{gray39}{rgb}{0.388235,0.388235,0.388235}
  \definecolor{grey39}{rgb}{0.388235,0.388235,0.388235}
  \definecolor{gray40}{rgb}{0.400000,0.400000,0.400000}
  \definecolor{grey40}{rgb}{0.400000,0.400000,0.400000}
  \definecolor{gray41}{rgb}{0.411765,0.411765,0.411765}
  \definecolor{grey41}{rgb}{0.411765,0.411765,0.411765}
  \definecolor{gray42}{rgb}{0.419608,0.419608,0.419608}
  \definecolor{grey42}{rgb}{0.419608,0.419608,0.419608}
  \definecolor{gray43}{rgb}{0.431373,0.431373,0.431373}
  \definecolor{grey43}{rgb}{0.431373,0.431373,0.431373}
  \definecolor{gray44}{rgb}{0.439216,0.439216,0.439216}
  \definecolor{grey44}{rgb}{0.439216,0.439216,0.439216}
  \definecolor{gray45}{rgb}{0.450980,0.450980,0.450980}
  \definecolor{grey45}{rgb}{0.450980,0.450980,0.450980}
  \definecolor{gray46}{rgb}{0.458824,0.458824,0.458824}
  \definecolor{grey46}{rgb}{0.458824,0.458824,0.458824}
  \definecolor{gray47}{rgb}{0.470588,0.470588,0.470588}
  \definecolor{grey47}{rgb}{0.470588,0.470588,0.470588}
  \definecolor{gray48}{rgb}{0.478431,0.478431,0.478431}
  \definecolor{grey48}{rgb}{0.478431,0.478431,0.478431}
  \definecolor{gray49}{rgb}{0.490196,0.490196,0.490196}
  \definecolor{grey49}{rgb}{0.490196,0.490196,0.490196}
  \definecolor{gray50}{rgb}{0.498039,0.498039,0.498039}
  \definecolor{grey50}{rgb}{0.498039,0.498039,0.498039}
  \definecolor{gray51}{rgb}{0.509804,0.509804,0.509804}
  \definecolor{grey51}{rgb}{0.509804,0.509804,0.509804}
  \definecolor{gray52}{rgb}{0.521569,0.521569,0.521569}
  \definecolor{grey52}{rgb}{0.521569,0.521569,0.521569}
  \definecolor{gray53}{rgb}{0.529412,0.529412,0.529412}
  \definecolor{grey53}{rgb}{0.529412,0.529412,0.529412}
  \definecolor{gray54}{rgb}{0.541176,0.541176,0.541176}
  \definecolor{grey54}{rgb}{0.541176,0.541176,0.541176}
  \definecolor{gray55}{rgb}{0.549020,0.549020,0.549020}
  \definecolor{grey55}{rgb}{0.549020,0.549020,0.549020}
  \definecolor{gray56}{rgb}{0.560784,0.560784,0.560784}
  \definecolor{grey56}{rgb}{0.560784,0.560784,0.560784}
  \definecolor{gray57}{rgb}{0.568627,0.568627,0.568627}
  \definecolor{grey57}{rgb}{0.568627,0.568627,0.568627}
  \definecolor{gray58}{rgb}{0.580392,0.580392,0.580392}
  \definecolor{grey58}{rgb}{0.580392,0.580392,0.580392}
  \definecolor{gray59}{rgb}{0.588235,0.588235,0.588235}
  \definecolor{grey59}{rgb}{0.588235,0.588235,0.588235}
  \definecolor{gray60}{rgb}{0.600000,0.600000,0.600000}
  \definecolor{grey60}{rgb}{0.600000,0.600000,0.600000}
  \definecolor{gray61}{rgb}{0.611765,0.611765,0.611765}
  \definecolor{grey61}{rgb}{0.611765,0.611765,0.611765}
  \definecolor{gray62}{rgb}{0.619608,0.619608,0.619608}
  \definecolor{grey62}{rgb}{0.619608,0.619608,0.619608}
  \definecolor{gray63}{rgb}{0.631373,0.631373,0.631373}
  \definecolor{grey63}{rgb}{0.631373,0.631373,0.631373}
  \definecolor{gray64}{rgb}{0.639216,0.639216,0.639216}
  \definecolor{grey64}{rgb}{0.639216,0.639216,0.639216}
  \definecolor{gray65}{rgb}{0.650980,0.650980,0.650980}
  \definecolor{grey65}{rgb}{0.650980,0.650980,0.650980}
  \definecolor{gray66}{rgb}{0.658824,0.658824,0.658824}
  \definecolor{grey66}{rgb}{0.658824,0.658824,0.658824}
  \definecolor{gray67}{rgb}{0.670588,0.670588,0.670588}
  \definecolor{grey67}{rgb}{0.670588,0.670588,0.670588}
  \definecolor{gray68}{rgb}{0.678431,0.678431,0.678431}
  \definecolor{grey68}{rgb}{0.678431,0.678431,0.678431}
  \definecolor{gray69}{rgb}{0.690196,0.690196,0.690196}
  \definecolor{grey69}{rgb}{0.690196,0.690196,0.690196}
  \definecolor{gray70}{rgb}{0.701961,0.701961,0.701961}
  \definecolor{grey70}{rgb}{0.701961,0.701961,0.701961}
  \definecolor{gray71}{rgb}{0.709804,0.709804,0.709804}
  \definecolor{grey71}{rgb}{0.709804,0.709804,0.709804}
  \definecolor{gray72}{rgb}{0.721569,0.721569,0.721569}
  \definecolor{grey72}{rgb}{0.721569,0.721569,0.721569}
  \definecolor{gray73}{rgb}{0.729412,0.729412,0.729412}
  \definecolor{grey73}{rgb}{0.729412,0.729412,0.729412}
  \definecolor{gray74}{rgb}{0.741176,0.741176,0.741176}
  \definecolor{grey74}{rgb}{0.741176,0.741176,0.741176}
  \definecolor{gray75}{rgb}{0.749020,0.749020,0.749020}
  \definecolor{grey75}{rgb}{0.749020,0.749020,0.749020}
  \definecolor{gray76}{rgb}{0.760784,0.760784,0.760784}
  \definecolor{grey76}{rgb}{0.760784,0.760784,0.760784}
  \definecolor{gray77}{rgb}{0.768627,0.768627,0.768627}
  \definecolor{grey77}{rgb}{0.768627,0.768627,0.768627}
  \definecolor{gray78}{rgb}{0.780392,0.780392,0.780392}
  \definecolor{grey78}{rgb}{0.780392,0.780392,0.780392}
  \definecolor{gray79}{rgb}{0.788235,0.788235,0.788235}
  \definecolor{grey79}{rgb}{0.788235,0.788235,0.788235}
  \definecolor{gray80}{rgb}{0.800000,0.800000,0.800000}
  \definecolor{grey80}{rgb}{0.800000,0.800000,0.800000}
  \definecolor{gray81}{rgb}{0.811765,0.811765,0.811765}
  \definecolor{grey81}{rgb}{0.811765,0.811765,0.811765}
  \definecolor{gray82}{rgb}{0.819608,0.819608,0.819608}
  \definecolor{grey82}{rgb}{0.819608,0.819608,0.819608}
  \definecolor{gray83}{rgb}{0.831373,0.831373,0.831373}
  \definecolor{grey83}{rgb}{0.831373,0.831373,0.831373}
  \definecolor{gray84}{rgb}{0.839216,0.839216,0.839216}
  \definecolor{grey84}{rgb}{0.839216,0.839216,0.839216}
  \definecolor{gray85}{rgb}{0.850980,0.850980,0.850980}
  \definecolor{grey85}{rgb}{0.850980,0.850980,0.850980}
  \definecolor{gray86}{rgb}{0.858824,0.858824,0.858824}
  \definecolor{grey86}{rgb}{0.858824,0.858824,0.858824}
  \definecolor{gray87}{rgb}{0.870588,0.870588,0.870588}
  \definecolor{grey87}{rgb}{0.870588,0.870588,0.870588}
  \definecolor{gray88}{rgb}{0.878431,0.878431,0.878431}
  \definecolor{grey88}{rgb}{0.878431,0.878431,0.878431}
  \definecolor{gray89}{rgb}{0.890196,0.890196,0.890196}
  \definecolor{grey89}{rgb}{0.890196,0.890196,0.890196}
  \definecolor{gray90}{rgb}{0.898039,0.898039,0.898039}
  \definecolor{grey90}{rgb}{0.898039,0.898039,0.898039}
  \definecolor{gray91}{rgb}{0.909804,0.909804,0.909804}
  \definecolor{grey91}{rgb}{0.909804,0.909804,0.909804}
  \definecolor{gray92}{rgb}{0.921569,0.921569,0.921569}
  \definecolor{grey92}{rgb}{0.921569,0.921569,0.921569}
  \definecolor{gray93}{rgb}{0.929412,0.929412,0.929412}
  \definecolor{grey93}{rgb}{0.929412,0.929412,0.929412}
  \definecolor{gray94}{rgb}{0.941176,0.941176,0.941176}
  \definecolor{grey94}{rgb}{0.941176,0.941176,0.941176}
  \definecolor{gray95}{rgb}{0.949020,0.949020,0.949020}
  \definecolor{grey95}{rgb}{0.949020,0.949020,0.949020}
  \definecolor{gray96}{rgb}{0.960784,0.960784,0.960784}
  \definecolor{grey96}{rgb}{0.960784,0.960784,0.960784}
  \definecolor{gray97}{rgb}{0.968627,0.968627,0.968627}
  \definecolor{grey97}{rgb}{0.968627,0.968627,0.968627}
  \definecolor{gray98}{rgb}{0.980392,0.980392,0.980392}
  \definecolor{grey98}{rgb}{0.980392,0.980392,0.980392}
  \definecolor{gray99}{rgb}{0.988235,0.988235,0.988235}
  \definecolor{grey99}{rgb}{0.988235,0.988235,0.988235}
  \definecolor{gray100}{rgb}{1.000000,1.000000,1.000000}
  \definecolor{grey100}{rgb}{1.000000,1.000000,1.000000}
  \definecolor{dark grey}{rgb}{0.662745,0.662745,0.662745}
  \definecolor{DarkGrey}{rgb}{0.662745,0.662745,0.662745}
  \definecolor{dark gray}{rgb}{0.662745,0.662745,0.662745}
  \definecolor{DarkGray}{rgb}{0.662745,0.662745,0.662745}
  \definecolor{dark blue}{rgb}{0.000000,0.000000,0.545098}
  \definecolor{DarkBlue}{rgb}{0.000000,0.000000,0.545098}
  \definecolor{dark cyan}{rgb}{0.000000,0.545098,0.545098}
  \definecolor{DarkCyan}{rgb}{0.000000,0.545098,0.545098}
  \definecolor{dark magenta}{rgb}{0.545098,0.000000,0.545098}
  \definecolor{DarkMagenta}{rgb}{0.545098,0.000000,0.545098}
  \definecolor{dark red}{rgb}{0.545098,0.000000,0.000000}
  \definecolor{DarkRed}{rgb}{0.545098,0.000000,0.000000}
  \definecolor{light green}{rgb}{0.564706,0.933333,0.564706}
  \definecolor{LightGreen}{rgb}{0.564706,0.933333,0.564706}